\documentclass[onecolumn,notitlepage,letterpaper,nofootinbib,longbibliography]{revtex4-2}

\usepackage[english]{babel}
\usepackage[T1]{fontenc}
\usepackage[utf8]{inputenc}
\usepackage{amsmath,accents}
\usepackage{amsfonts,color}
\usepackage{amssymb}
\usepackage{float}
\usepackage{ragged2e}
\usepackage{comment}
\usepackage{mathrsfs}
\usepackage{xurl}
\usepackage[margin=0.6in]{geometry}
\usepackage{graphicx}
\usepackage{caption}
\usepackage{ragged2e} 
\usepackage[bottom]{footmisc}
\usepackage{comment}

\usepackage{tikz-cd}

\usepackage{graphicx}

\usepackage[pdftoolbar = true, pdfstartview = FitH, pdfmenubar = true]{hyperref}
\hypersetup{
    colorlinks=true,
    linkcolor=blue,
    filecolor=magenta,      
   citecolor=blue
}

\begin{document}

\title{Quasi-Palatini Formulation of Scalar-Tensor Gravity}

\author{Sotirios Karamitsos}
\email{sotirios.karamitsos@ut.ee}
\affiliation{Institute of Physics, University of Tartu, W.\ Ostwaldi 1, 50411 Tartu, Estonia}

% Keywords
%\keyword{Scalar-tensor gravity, inflation, dynamical systems}

\begin{abstract}
The Palatini formulation has been successful in the development of several alternative theories of gravity. It is well understood that the Palatini and metric formulations are equivalent in minimally coupled scalar-tensor models, but nonminimal scalar-tensor models can lead to physically distinct theories depending on the underlying formulation. Once a model has been selected, the choice of formulation is a discrete one, and so promoting it to be continuous is expected to give rise to a wider class of actions. To this end, we propose the ''quasi-Palatini'' formulation, a method for interpolating between the metric and Palatini formulations for a given model that gives rise to a continuous family of models. We apply the quasi-Palatini formulation to Higgs inflation, induced gravity inflation, and Starobinsky inflation, and demonstrate how this leads to a deformation of the potential, studying its impact on observables. We also discuss how the interpolation between different actions can be extended to scalar-torsion and scalar-nonmetricity models.
\end{abstract}

\maketitle
\section{Introduction}

Inflation has become a fixture of standard cosmology, owing in no small part to the elegant solutions it offers to the flatness, horizon, and relic problems of Big Bang cosmology. Moreover, inflation further provides a generic mechanism by which the large scale structure observed in the universe today can be traced all the way back to the evolution of the primordial perturbations. At the same time, a very large number of models has been proposed, with many of them not surviving the ever-tightening observational window, which has motivated the development of more sophisticated models. 

In order for inflation to happen, either the gravitational sector has to be modified, or some additional matter has to be introduced. A popular mechanism through which inflation can be driven is the wide class of scalar-tensor gravity. Theories in this class are specified by a number of model functions which act as parameters specifying the model, such as the non-minimal coupling, the kinetic coupling, and the potential. Scalar-tensor theories of gravity may be expressed in different frames through a conformal transformation and a field reparametrization. There has been much discussion on the physical meaning of frame transformations, but the consensus is that classically, different frames of the same theory ultimately carry the same physical content \cite{Flanagan:2004bz,Chiba:2013mha,Postma:2014vaa,Burns:2016ric,Jarv:2016sow,Karamitsos:2017elm}, which may be  may hold at the quantum level as well \cite{Hill:2020oaj}, as a result of the fact that such transformations are akin to changes of units
\cite{Dicke:1961gz,Faraoni:1998qx}. 

The existence of different cosmological frames is just one example of equivalent representations of the same underlying theory. For instance, there are multiple variational principles that can be employed to derive the Einstein field equations. One of the earliest and most enduring is the so-called Palatini variation, actually studied by Einstein himself originally~\cite{Ferraris:1982wci}. In the Palatini approach, there is no \emph{a priori} assumption as to the form of the connection, and the variation of the Einstein-Hilbert action with respect to the connection imposes the Levi--Civita form, which is imposed ``by hand'' in the usual metric approach. The Palatini variation is an example of the more general notion of metric-affine gravity, in which the notion of parallel transport is also decoupled from the notion of geodesics. 

The Palatini approach is often termed a \emph{formulation} rather than a theory. This is appropriate, since the Palatini variation does not yield additional degrees of freedom in General Relativity (GR). The same holds for the alternative teleparallel equivalents of GR \cite{Nester:1998mp,Maluf:2013gaa,Mol:2014ooa} that together with GR form the so-called geometric trinity of gravity~\cite{BeltranJimenez:2019esp}. Therefore, we have a choice between different actions that yield the same dynamics: the same degrees of freedom are encoded in a different way. This is precisely why it is apt to call Palatini and the teleparallel equivalents of GR different formulations of the same \emph{theory}: if their physical content did not match, as is the case for more general metric-affine gravity models, then they would constitute \emph{bona fide} distinct theories rather than different formulations of the same theory.

In the case of modified gravity with a nonminimal coupling, however, the situation is different. Following the same procedure that we use to construct the actions that correspond to different formulations to GR, but with a nonminimal action as a starting point, we arrive at an action that no longer carries the same physical content to the original. This makes ``formulation'' somewhat of a misnomer in this case, but the term has persisted. As such, in practice throughout the literature, taking an established action and ``switching between formulations'' is used to motivate different theories, such as Palatini scalar-tensor gravity \cite{Bauer:2008zj}, Palatini $F(R)$ gravity \cite{Olmo:2011uz}, Palatini $F(R,T)$ gravity \cite{Wu:2018idg}, with applications to cosmological perturbations \cite{Koivisto:2005yc} as well as specific models such as Palatini Higgs inflation \cite{Bauer:2010jg}, Palatini attractors \cite{Jarv:2017azx}, and quintessential $R^2$ inflation \cite{Dimopoulos:2022rdp}.

The choice of formulation is generally thought of as discrete: once a model has been decided upon, we can choose to work with metric or the Palatini formulation, leading to different physics. This invites the question of whether we can promote this discrete choice to a continuous one. To be more precise, since choosing between the two formulations leads to two physically distinct theories, we would expect that interpolating between the two formulations would lead to a continuous class of physically distinct theories, indexed by a single parameter. For scalar-tensor theories or $F(R)$ theories (studied in their equivalent scalar-tensor form), we expect the resulting interpolated theories to also be scalar-tensor. This is because ``Palatini scalar-tensor gravity'' does not in itself constitute a distinct class of theories, since the Palatini formulation does not ``broaden'' the space of scalar-tensor actions, as it leaves their form invariant. Therefore, a way to interpolate between the two formulations would offer a novel method to motivate several models that can be matched against observations. 

It would be a glaring omission here not to mention the so-called \emph{hybrid} metric-Palatini theories \cite{Harko:2011nh,Koivisto:2013kwa,Capozziello:2015lza}. Such theories feature a modification of the conventional Einstein--Hilbert action by the addition of  metric-affine (i.e. Palatini) curvature terms. This approach can help unify local tests at the Solar System level with the late-time cosmic acceleration, even if the scalar field is very light. 
%Hybrid metric-Palatini theories have been studied both in the context of modified gravity, including $f(R)$ theories, as well as Higgs inflation. 
%
However, hybrid metric-Palatini theories, despite their many advantages, do not constitute (a class of) formulations as the term is commonly understood: they bridge together the distinct classes of metric and Palatini actions, but they do not interpolate between metric and Palatini realizations of the \emph{same} model. As such, despite being a fruitful approach to model building, they are best described as a \emph{deformation} of GR rather than a formulation.

It is reasonable to ask that any interpolation between the metric and Palatini formulations does not introduce any new functions, just as is the case when switching between formulations. Therefore, to this end, we propose the  propose the \emph{quasi-Palatini}\footnote{The author also considered ``semi-Palatini'', ``partial Palatini'' or ``mixed Palatini'' before eventually settling on ``quasi-Palatini''.} formulation, in which the binary choice between the two formulations is replaced by a single parameter which controls the relative ``strength'' of metric or Palatini: once an model has been decided upon, it can be studied in ``$50\%$ metric'' or ``$25\%$ Palatini'' formulation in addition to ``pure metric'' or ``pure Palatini''. Each of these formulations leads to a distinct theory with distinct predictions. The same philosophy can be applied to the teleparallel formulation of gravity, making it possible to express different models in the ``quasi-teleparallel'' formulation, again leading to manifestly different theories (though belonging to a richer class than scalar-tensor theories owing to non-vanishing boundary terms).  

The structure of this article is as follows. In Section~\ref{metricvsPalatini}, we outline the distinction between the metric and Palatini formulations, and examine the effects of conformal transformations on scalar-tensor gravity depending on the choice of formulation. In Section~\ref{actionstheories}, we draw a distinction between models, actions, and theories, and briefly overview the different formulations in the geometric trinity of gravity before formalizing the notion of a formulation of a model. In Section~\ref{quasi}, we motivate the quasi-Palatini formulation and demonstrate how it differs from hybrid metric-Palatini gravity. We discuss how to interpolate between actions of models interpreted in different formulations, and study scalar-tensor models and $F(R)$ models in the quasi-Palatini formulation before discussing how this interpolation of formulations may be extended to teleparallel gravity. In Section~\ref{infl}, we study well-known inflationary models such as Higgs inflation, induced gravity inflation, and Starobinsky inflation in the quasi-Palatini formulation, demonstrating how the interpolation between different formulations can be used to define a continuous class of theories without additional scalar degrees of freedom. Finally, we discuss our findings and conclude in \ref{concl}.

Throughout this paper, we work in natural units in which $\kappa^2 = 8\pi G = M_P^{-2} = 1$, and we employ the predominantly plus convention for the signature of the metric. 

\section{The metric and Palatini formulations}
\label{metricvsPalatini}

\subsection{Frame transformations in metric and Palatini} 
Early on in the history of the study of gravity, it was understood that variational principles would play an important role in the study not only of GR, but its modifications as well. Using the Einstein--Hilbert action as a starting point, it is possible, depending on the context and motivation, to modify the gravity sector, add a matter sector, or possibly both, leading to a host of actions with different predictions. The Einstein--Hilbert action with a nonminimally coupled matter sector is given by
\begin{align}\label{ehaction}
S_{\rm metric}[g_{\mu\nu},\phi] = \frac{1}{2} \int {\rm d}^4 x \, \sqrt{-g} \,   f(\phi) R +  S_m[g_{\mu\nu},\phi] ,
\end{align}
where the scalar curvature $R = R(g_{\mu\nu})$ is taken to be a function of the metric by assuming that the connection~$\Gamma^\rho_{\mu\nu}$ is the torsionless and metric-compatible Levi--Civita connection, and $\phi$ is a stand-in for all matter fields (including their derivatives). We will leave a more careful treatment of $f(R)$ theories aside for now, but we note that $f(R)$ theories can be recast as scalar-tensor theories through a Legendre transformation of the action, making the following discussion applicable to them as well.

The idea that the connection can be independent of the metric was explored early on in the history of GR. This approach was further refined into metric-affine gravity  \cite{Vitagliano:2010sr}, in which the connection need be neither torsionless nor metric-compatible. This is essentially a decoupling of the notions of geodesics and parallel transport. Therefore, in a way, the metric-affine approach is more ``natural'' than the pure metric one \cite{El1973} as it is not necessary for the connection to take on the Levi--Civita form by hand. 
%Furthermore, metric-affine gravity is more amenable to generalizations, and may even be easier to quantize \cite{}. 

The Palatini formulation is a special case of metric-affine gravity, with the difference that, as opposed to the general case, the independent connection does not couple to matter.
Therefore, the action, which is now a functional not only of~$g_{\mu\nu}$, but of~$\Gamma^\rho_{\mu\nu}$ as well, takes on the following form:
\begin{align}\label{palatiniaction}
S_{\rm Palatini}[g_{\mu\nu},\Gamma^\rho_{\mu\nu},\phi] = \frac{1}{2} \int {\rm d}^4 x \, \sqrt{-g} \,  f(\phi) \mathcal{R} +  S_m[g_{\mu\nu},\phi].
\end{align}
We will focus on first-order gravity for now and consider higher order gravity later. Here $\mathcal{R}$ is the \emph{generalized} Ricci scalar. The difference between $R$ and $\mathcal{R}$ is that, even though they both have the same formal appearance, $\mathcal{R}$ is a function of both $g_{\mu\nu}$ and $\Gamma^\rho_{\mu\nu}$, whereas $R$ is a function of $g_{\mu\nu}$ by imposing the Levi-Civita form on the connection~$\Gamma^\rho_{\mu\nu}$. Written with all the arguments made explicit:
\begin{align}\label{rdef}
R(g)&\equiv  g^{\mu\nu}R_{\mu\nu}(g)  \: \equiv g^{\mu\nu} \left(   {}^{(g)}\Gamma^\rho_{\mu \nu,\rho} -  {}^{(g)}\Gamma^\rho_{\nu \rho,\mu} + {}^{(g)}\Gamma^\rho_{\rho \lambda} {}^{(g)}\Gamma^\lambda_{\mu \nu} - {}^{(g)}\Gamma^\rho_{\mu \lambda} {}^{(g)}\Gamma^\lambda_{\nu \rho}\right),
\\
\label{calrdef}
\mathcal{R}(g, \Gamma) &\equiv  g^{\mu\nu}\mathcal{R}_{\mu\nu}(\Gamma)   \equiv  g^{\mu\nu} \left(   \Gamma^\rho_{\mu \nu,\rho} -  \Gamma^\rho_{\nu \rho ,\mu} + \Gamma^\rho_{\rho \lambda} \Gamma^\lambda_{\mu \nu} - \Gamma^\rho_{\mu \lambda} \Gamma^\lambda_{\nu \rho}\right),
\end{align}
where $R_{\mu\nu}$ is the usual Ricci tensor, defined in terms of $g_{\mu\nu}$ through the Levi-Civita connection~${}^{(g)}\Gamma^\rho_{\mu\nu}$ compatible with $g_{\mu\nu}$, and $\mathcal{R}_{\mu\nu}$ is the similarly generalized Ricci tensor, defined solely in terms of the indepenent connection $\Gamma^\rho_{\mu\nu}$ whose form is not assumed.

It is well known that the equations of motion of \eqref{ehaction} and \eqref{palatiniaction} coincide when $f(\phi)$ is a constant, i.e. when both actions are minimally coupled. However, in the presence of a nonminimal coupling, subtleties arise regarding the relation between the metric and the independent connection. Extremizing \eqref{ehaction} with respect to the metric yields the usual Einstein field equations, but in case of the Palatini action \eqref{palatiniaction}, we must extremize it both with respect to the metric and the connection. The resulting equations of motion are as follows: first, from varying with respect to the metric,
\begin{align}
f(\phi) \mathcal{R}_{(\mu\nu)} - \frac{f(\phi)}{2} = T_{\mu\nu},
\end{align}
where as usual the energy-momentum tensor is $T_{\mu\nu}\equiv -\frac{2}{\sqrt{-g}} \frac{\delta S_m}{\delta g^{\mu\nu}}$, and second, from varying with respect to the torsion-free connection $\Gamma$, 
\begin{align}\label{connectioneom}
\nabla^{\Gamma}_\rho ( \sqrt{-g} f(\phi) g^{\mu\nu}) &= 0.
\end{align}
There is no reason to assume from the start that the connection is torsionless, but it can be shown that that, if present, the torsion is necessarily sourced by a single vector, which means that it does not enter the equations of motion, and therefore the common practice of setting it to zero in the Palatini formulation from the start is justified \cite{Olmo:2011uz}. 

It is common to say that $R$ and $\mathcal{R}$ are conformally related due to \eqref{connectioneom}, but this statement is somewhat misleading.  For a conformal transformation\footnote{More appropriately termed a \emph{Weyl transformation}, since it is a local rescaling of the metric rather than a coordinate transformation, but calling it a conformal transformation prevails in the literature.} given by
\begin{align}\label{conftrans}
 g_{\mu\nu} \to \hat  g_{\mu\nu} = \Omega^2 g_{\mu\nu},
\end{align}
we can always write the following identities that link $R$ and $\mathcal{R}$ through their definitions \eqref{rdef} and \eqref{calrdef}:
\begin{align}
\label{idrelRR}
\mathcal{R}(g, {}^{(g)}\Gamma) &\equiv R(g),
\\
\label{idrelRR2}
\mathcal{R} ( g, {}^{(\hat g)}\Gamma^\rho_{\mu\nu}) &\equiv \Omega^{ 2}\mathcal{R}( \hat g, {}^{(\hat g)}\Gamma^\rho_{\mu\nu})
\end{align}
These identities always hold. However, if we want to relate $R$ and $\mathcal{R}$ on-shell, we must necessarily first determine the relationship between the independent connection $\Gamma^\rho_{\mu\nu}$ and the metric $g_{\mu\nu}$: only then can we write the relation between $R$ and $\mathcal{R}$.

The transformation rules for $R$ and $\mathcal{R}$ under a conformal transformation \eqref{conftrans} are:
\begin{align}
\label{transf1}
R(g) \to \  R(\hat g) &= \Omega^{-2} \left[R(g) - 6\Omega^{-1} \nabla^2 \Omega\right],
\\
\label{transf2}
\mathcal{R}(g,\Gamma) \to  \mathcal{R}(\hat g, \Gamma) &= \Omega^{-2} \mathcal{R}(g,\Gamma).
\end{align}
As usual, the rule for $R$ has a derivative term thanks to the Levi--Civita connection transforming along with the metric, whereas $\mathcal{R}$ only picks up a prefactor, since the independent connection does not transform. Now, through \eqref{connectioneom}, we can write the explicit form of the connection as
\begin{align} \label{connsol}
 \Gamma^\rho_{\mu\nu} = {}^{(\hat g)}\Gamma^\rho_{\mu\nu},
\end{align}
for the particular choice of $\Omega^2 = f(\phi)$. Then, combining the expressions \eqref{idrelRR} and \eqref{idrelRR2}, we can write $\mathcal{R}(g, {}^{(\hat g)}\Gamma)  \equiv f^{-2} R(\hat g)$, and using the conformal transformation rule for $R$ given in \eqref{transf1}, we get
\begin{align}\label{relationrR}
\mathcal{R} &\equiv R  - 6 f^{-1/2} \nabla^2 \sqrt{f},
\end{align}
where $\mathcal{R} = \mathcal{R}(g, {}^{(\hat g)}\Gamma)$ and $R = R(g)$. Therefore, the relation between $R$ and $\mathcal{R}$ is similar to but does not match the conformal transformation rule: the precise statement is that on-shell, the independent connection (through which $\mathcal{R}$ is defined) takes on the Levi-Civita form compatible with the conformally related metric $f(\phi)g_{\mu\nu}$. 
%The relation \eqref{relationrR} is analogous to the relation found in hybrid metric-Palatini theories in which $\mathcal{R}$ is expressed algebraically in terms of $X \equiv R + T^\mu_{\ \mu}$, where $X$ measures the deviation from the GR trace equation~\cite{Capozziello:2015lza}.

It is worth restating that the particular ``conformal'' relation between ${R}$ and $\cal R$~\eqref{relationrR} only holds thanks to the validity of the particular solution~\eqref{connsol}. The conformal transformation \eqref{conftrans} followed by an appropriate reparametrization $\phi \to \varphi = \varphi(\phi)$ can be used to transform the action between the frames frames. Therefore, it is only meaningful to relate $\mathcal{R}$ and $R$ \emph{after} a frame has been chosen and the exact form of the action has been specified. Thus, it a \emph{frame-dependent} statement: working in a frame where $f(\phi)$ is a constant means that $R$ and $\cal R$ match, but this does not hold in any other frame. Nonetheless, the existence of this relation has the fortunate effect that such actions in Palatini gravity can be recast solely in terms of the metric (and as such the usual Ricci scalar).

\subsection{Scalar-tensor theories}

We further specialize to the rather wide and versatile class of scalar-tensor theories, which are described by the three model functions: the nonminimal coupling $f(\phi)$ (which couples to the Ricci scalar as described) the kinetic coupling matrix $k_{AB}(\phi)$, and the potential $V(\phi)$:
\begin{align}\label{scaltensact}
S^{\rm (ST)}  &=  \int {\rm d}^4 x \, \sqrt{-g} \,   \left[ \frac{f(\phi)R}{2} - \frac{k_{AB}(\phi)}{2} (\partial_\mu\phi^A) (\partial^\mu\phi^A) - V(\phi) + {\cal L}_m\right]  ,
\end{align}
where $\phi$ now stands for an array of fields $\phi^i$ when appearing as an argument. While we will focus on single-field models for most of this paper, using the more general multifield action here will make the geometric interpretation of the action more manifest.

%Scalar-tensor theories have been studied extensively, particularly with respect to their behavior under conformal transformations. Even if there is still some ambiguity regarding the physical meaning of conformal transformations after quantization \cite{theframestuff}, it is now generally understood that at the classical level, the effects of conformal reparametrizations amount to no more than a reshuffling of the degrees of freedom.

The fact that the frame in which a theory is expressed has no physical meaning classically (i.e. there is no experiment that can distinguish between which frame we live in) can be distilled to the observation that $f(\phi)$ is not a physical quantity. This can be understood as $f(\phi)$ essentially setting the system of units, which, even though spacetime-dependent through $\phi$, cannot be observed, which is also confirmed the fact that the conformally covariant derivative of $f(\phi)$ (defined by way of the Weyl derivative) which measures the physically meaningful variation of a quantity up to a change of frame \cite{Postma:2014vaa,Karamitsos:2017elm}, vanishes:
\begin{align}
\mathcal{D}_\mu X = \partial_\mu f - w_X \ln N_L
\end{align}
where $w_X$ is the conformal weight of the quantity $X$ and $N_L$ is the lapse function with $d_{N_L} = -1$ since $\Omega \to \Omega N_L$. We can readily see that $\mathcal{D}_\mu f = 0$ since $w_f = -2$.

The discussion above does not imply that modifying $f(\phi)$ does not change the physical content of the theory. Indeed, if we ``index'' the action \eqref{scaltensact} above as $S^{\rm (ST)}_m[f(\phi),k_{AB}(\phi),V(\phi), {\cal L}_m ]$, then, given a function $\tilde f(\phi)$, there is exactly one action $S^{\rm (ST)}_m[\tilde f(\phi), \tilde k_{AB}(\phi), \widetilde V(\phi),\widetilde {\cal L}_m ]$ that is completely equivalent and cannot be distinguished via observations. 

Switching to the Palatini interpretation means replacing $R$ with $\cal R$, making the action
\begin{align}\label{scaltensactPAL}
S^{\rm (ST)}  &=  \int {\rm d}^4 x \, \sqrt{-g} \,   \left[ \frac{f(\phi){\cal R}}{2} - \frac{k_{AB}(\phi)}{2} (\partial_\mu\phi^A) (\partial^\mu\phi^A) - V(\phi) + {\cal L}_m\right].
\end{align}
We may transform such an action to its metric analogue through \eqref{relationrR} before making a conformal transformation \eqref{conftrans}. This transformation leaves the action form invariant, though the transformation rules for the model functions differ depending on whether we started from the metric or Palatini action. We have
\begin{align}\label{ct1}
\tilde f &= \Omega^{-2} f
\\ \label{ct2}
\tilde k_{AB} &=  \Omega^{-2} \left[k_{AB} - 6 \delta_m f  (\ln \Omega)_{,A} (\ln\Omega)_{,B} + 6 \delta_m f_{,(A} (\ln \Omega)_{,B)}\right],
\\ \label{ct3}
\widetilde V &= \Omega^{-4} V,
\\
\widetilde {\cal L}_m &= \Omega^{-4} {\cal L}_m,
\end{align}
where for convenience, we define
\begin{align}
\delta_m \equiv 
\begin{cases}
1 & {\rm (metric)}
\\
0 & {\rm (Palatini)}
\end{cases} \quad,
\qquad
\delta_P \equiv 
\begin{cases}
0 & {\rm (metric)}
\\
1 & {\rm (Palatini)}
\end{cases} \quad .
\end{align}
After the conformal transformation to the Einstein frame (with $\Omega^2 = f$, i.e. choosing $\tilde f = 1$), the kinetic term takes on the form
\begin{align}\label{fieldmetric}
G^{(P)}_{AB}(\phi) &= \frac{k_{AB}}{f} + \frac{3\delta_m}{2} \frac{f_{,A}f_{,B}}{f^2}. 
\end{align}
and the potential takes on the term 
\begin{align}
V_E(\phi) = \frac{V(\phi)}{f(\phi)^2}.
\end{align}
The difference between the form of the kinetic term between metric and Palatini follows from the transformation rules \eqref{transf1} and \eqref{transf2}, and depends whether we take the fields to couple between the Ricci scalar (metric) or the generalized Ricci scalar (Palatini).

\begin{figure}[t]
     \includegraphics[width=0.60\linewidth]{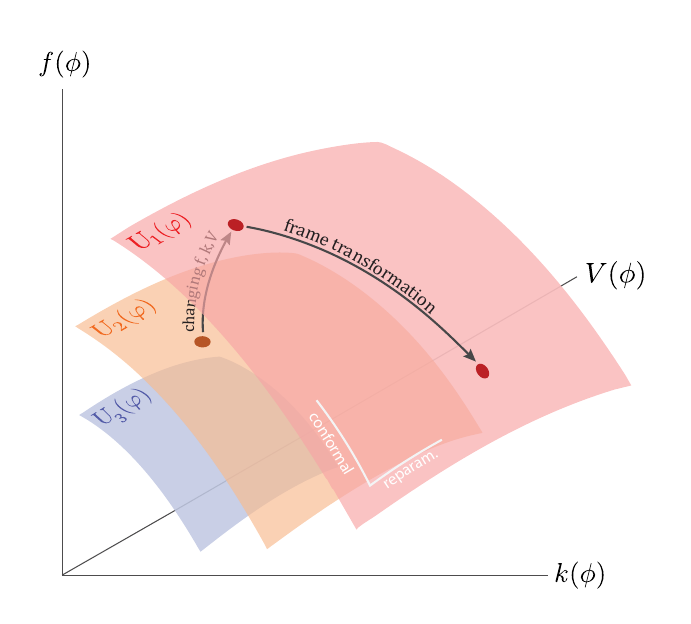} 
    \caption{\justifying 
    The space of single-field scalar-tensor metric actions, parametrized by model functions. Points in this space correspond to different actions. Actions on the same shell have the same invariant potential, and so belong to the same equivalence class, i.e. they contain the same physics, just expressed in different frames. Conformal transformations and field reparametrizations do not change the invariant potential, and so frame transformations keep the actions in the same frame equivalence class, whereas an unconstrained change in the functions $f(\varphi)$, $k(\phi)$, and $V(\phi)$ that does not satisfy \eqref{ct1}-\eqref{ct3} takes actions outside the equivalence class. 
    %Note that $F(R)$ theories can be represented (after a Legendre transform) by actions living on a line in the $f-V$ plane, since for them, $k(\phi) = 0$ and there is an algebraic relation between $f(\phi)$ and $V(\phi)$ that fully specifies them (and after a conformal transformation
    }
    \label{fig:modelsframe}
\end{figure}

Regardless of whether we are working in metric or Palatini, interpreting $G_{AB}$ as metric gives rise to the field space, which is a Riemannian manifold (if the kinetic term is positive-definite to avoid ghosts)s. In particular, we can always choose to work in polar coordinates. If we define a particular transformation $\phi \to \varphi = \varphi (\phi)$ with the associated Jacobian
\begin{align}
J^A_{\ B} \equiv \frac{d\varphi^A}{d\phi^B},
\end{align}
the line element can be written as follows:
\begin{align}\label{polarkinetic}
d\sigma^2 = d\rho^2 +  k_1 (\rho,\varphi) (d\varphi^1)^2 + \ldots+ k_{N-1}(\rho,\varphi)(d\varphi^{N-1})^2,
\end{align}
where $\varphi^A = (\rho, \varphi^a)$, and $a$ runs over $(1,\ldots, N-1)$, where $N$ is the number of fields. To avoid any ambiguity, we note that we define the Jordan frame as any frame where nonminimal coupling $f(\phi)$ is not a constant (as usual), the Einstein frame as any frame in which $f(\phi)$ is a constant, and the canonical Einstein frame as the \emph{single} frame where the kinetic terms take on the polar form given by \eqref{polarkinetic}, i.e. at least one field has a kinetic prefactor equal to 1 (possibly more depending on the spectral structure of $k_{AB}$), and there are no cross-terms (although in practice it may be very cumbersome to write in closed form in terms of the original fields).

Generalized invariants for scalar-tensor theories have been explicitly calculated in \cite{Jarv:2014hma, Kuusk:2015dda}. A scalar-tensor theory with~$N$ fields can be specified by choosing $N(N+1)/2 +2$ functions (one nonminimal coupling $f(\phi)$, $N(N+1)/2$ quadratic kinetic terms $k_{AB}$, and one potential $V(\phi)$). However, working in the Einstein frame and radially parametrizing the fields via a polar chart shows that there are $N$ functional degrees of freedom in this class of theories: $N-1$ diagonal kinetic terms plus the invariant potential $U(\varphi)$, and as such, $N$ independent invariant quantities. For a single field, we find the usual result that all the complexities of introducing a nonminimal coupling $f(\phi)$ or a nontrivial kinetic coupling $k(\phi)$ can be absorbed into the choice of a single function $U(\varphi)$, which corresponds to the potential in the Einstein frame.\footnote{Subtleties arise when the nonminimal coupling or kinetic coupling is singular, such as in $\alpha$-attractors. Such actions harbor domain walls, crossing which is impossible due to the infinite field space distance between points on either side~\cite{Karamitsos:2019vor, Karamitsos:2021mtb}. Therefore, such action can be mapped to multiple minimal actions (the number of which depends on the precise structure of the kinetic matrix), all of which still have $N$ functional degrees of freedom.}

We can schematically illustrate the physical and nonphysical components of frame transformations (i.e. conformal transformations and field reparametrizations) for $N=1$ in Fig.~\ref{fig:modelsframe}. The precise form of the frame transformation rules depends on the initial choice between metric and Palatini, but the overall structure of the flow between points in the abstract ``action space'' remains the same. We remind that there is no need to differentiate between the metric and Palatini scalar-tensor action spaces, since it is always possible to relate the Ricci scalar and the generalized Ricci scalar for a given $f(\phi)$. All $F(R)$ theories can also be embedded in the metric scalar-tensor action space, since a Legendre transformation can be used to write an equivalent scalar-tensor action (and since we only have one functional degree of freedom in writing $F(R)$ theories, they correspond to a single line in action space, each point of which corresponds to a different invariant potential). The effects of conformal transformations and field reparametrizations on an arbitrary action correspond to transforming between points in an abstract ``action space'', which can be partitioned into equivalence classes up to their invariant potential. For $N>1$, the equivalence class surfaces become~$(N(N-1)/2+2)$-dimensional hypersurfaces embedded in the~$(N(N+1)/2+2$-dimensional space of $N$-field multiscalar-tensor theories. We will formalize this idea in the next section.

\section{Actions, theories, models, and formulations}
\label{actionstheories}

The starting point for every modified theory of gravity is its action. Once an action is written down, then the physical content of the theory it corresponds to is uniquely specified, and the theory can be studied. With so many different modifications to gravity, each with its own action, it is only natural to ask whether some of these contain the exact same physics, just in a different package. After all, ``the map is not the territory'', and just like there are multiple projections in map-making, there are multiple equivalent actions that describe the same physics. Still, ``action'' and ``theory'' are two terms that are very often used interchangeably. In isolation, this is not problematic: after all, a well-specified action unambiguously defines a theory. As we shall see, however, for a reasonable definition of ``theory'', the converse does not hold, and so it will be helpful to formally understand the relation between the two before moving forward.

\subsection{The theory space as the quotient of the action space}

We mentioned the action space informally in the previous section, and now it is time to make an attempt to formalize it. We can identify the space of metric scalar-tensor actions $\mathbb{A}^{(m)}_{\rm ST}$ as a topological vector space.  Plainly put, fixing $f$, the independent components of $k_{AB}$, and $V$ is enough to unambiguously specify a scalar-tensor action, Similarly, we may  identify the space of Palatini scalar-tensor actions $\mathbb{A}^{(P)}_{\rm ST}$, where once again fixing the model parameters unambiguously defines an action. These belong to the huge space of functionals, and particularly the smaller (but still immense) space of functionals that make sense as actions (nondimensionalizable, free from instabilities, and taking continuous functions as arguments among other requirements).

An action space does not have an inherent geometrical structure: it is conceptually distinct from the configuration space in which systems evolve. Instead, a point in action space has an entire configuration space associated with it: once an action is picked, we can then study the evolution of the system that it describes. At the simplest level, a (real) action space is a direct product of countably multiple copies of the topological space of (smooth, for our purposes) functions $C^\infty(\mathbb{R})$, and as such it is a Fr\'echet space. This is a complete topological vector space that does not admit a natural topology, i.e. even though its topology can be defined by a a countable family of norms (corresponding loosely to the multiple functional ``directions''), there is no natural topology induced by a continuous norm-induced metric, even though an intuitive notion of closeness still exists.

Within an action space, we can view frame transformations as defining a continuous group action\footnote{Unfortunately, the term ``action'' pulls double duty here. We will always use the full term \emph{group action} to avoid any ambiguities.}. A frame transformation can be unambiguously specified by the conformal factor and the related rescaling, and as such the associated group is ${\cal F} =\mathbb{R}^+_\times \otimes GL_N(\mathbb{R})$. We can also define an equivalence relation $\sim$ between actions in the same space that can be transformed into one another following a frame transformation. This relation gives rise to equivalence classes: these can also be motivated as induced by the orbit of the group action. For scalar-tensor theories, these correspond to the sheets of~Fig.~\ref{fig:modelsframe}. The group action inherits the usual features of frame transformations: for instance, the group action is not transitive (since there are actions that cannot be transformed to one another), and the little group of a given action corresponds to $\Omega = 1$ and the cases where $J^A{}_B$ is a permutation matrix (in which case the action under the group action is mapped to itself). 

The formal understanding of what makes two actions ``physically equivalent'' can be gleaned through the notion of physical (``engineering) dimensions, and their extension to conformal dimensions. Given an action, it is possible to extract physical laws (e.g. the Einstein field equations) and predictions (e.g. the tilt of the spectrum, the tensor-to-scalar ratio), which may well be an arduous process. However, it is an unavoidable fact that every such relation can always be written in terms of frame-invariant quantities. This is essentially a consequence of the Buckingham-$\pi$ theorem  \cite{buckingham1914}, which plainly states that choice of units cannot be physical, and the understanding of frame transformations as a spacetime-dependent redefinition of units \cite{Dicke:1961gz}. Of course, we can write an infinite set of invariant quantities, each with a potentially different physical interpretation, but the minimal set of $\pi_i$ independent, frame-invariant quantities that suffice to write all nondimensionalized, physically meaningful relations $f_n (\pi_j) = 0 $ of the theory forms a basis (one of many). Therefore, if there is a change of basis between two sets of relations $f_n (\pi_j) = 0 $ and $f'_n (\pi'_j) = 0$ such that $f_n (a_{ij} \pi_j) = f'_n (\pi'_i)$, these two sets encode the same physics. 

We can therefore decouple the notion of a scalar-tensor \emph{theory} from its multiple \emph{representations} (i.e. actions) by identifying the \emph{theory space} of scalar-tensor gravity $\mathbb{T}_{\rm ST}$ as the quotient of the action space under frame transformations, or
\begin{align}
\mathbb{T}_{\rm ST} \equiv \mathbb{A}_{\rm ST}/{\cal F}.
\end{align}
The theory space of scalar-tensor actions is isomorphic to the direct product of $N$ copies of $C^\infty(\mathbb{R})$. This is just the formalization of the fact that such a theory can be indexed by $N$ arbitrary functions (and the most convenient way to parametrize these is such that they return the action in its polar form). For single-field actions, the theory space is just the space of continuous functions, indexed by the invariant potential $U(\varphi)$. A theory then is nothing more than an equivalence class of actions.

There are more transformations that do not change the physical content of a theory beyond just frame transformations.  For instance, we can consider the action space of $F(R)$ theories ${\cal A}_{F(R)}$, defined by a single functional parameter $F$. Via the Legendre transform, there is a class of meromorphisms from ${\cal A}_{F(R)}$ to ${\cal A}_{\rm ST}$ (indexed by the arbitrary value of the nonminimal coupling, although in the common Brans--Dicke form this is usually taken to be $\phi$, and the morphism is non-surjective since there are actions that do not correspond to an $F(R)$ theory), and there is an isomorphism between $\mathbb{A}_{F(R)}$ and $\mathbb{T}_{F(R)}$ (for a single field), implying that $F(R)$ theories (for well-behaved functions $F$) are no more ``richer'' than scalar-tensor theories \cite{Capozziello:2010ih}.

Even though $\mathbb{A}^{\rm metric}_{\rm ST}$ and $\mathbb{A}^{\rm Palatini}_{\rm ST}$ are indeed distinct action spaces, they are isomorphic through \eqref{relationrR}. This essentially says that, given model parameters $f$, $k_{AB}$ and $V$, coupling to the generalized Ricci scalar instead of the usual metric Ricci does not ``enrich'' the theory space. There are many good practical reasons why we may choose to study a particular representation of a \emph{theory} (in the Einstein or Jordan frame, cast as a $F(R)$ action, in the metric or Palatini formulations), which depend on how the action is motivated, but the distinction between actions and theories remains, all coming down to the simple fact that multiple actions can encode the same physics, and the same theory can be expressed in multiple ways.

The $G$-invariants of the action spaces are the minimal actions living on the slice $f'(\phi) = 0$. The induced equivalence classes are either singletons (the minimal actions) or pairs of nonminimal actions that map to each other after the switch, which means that the groupoid action (defined on the subset of nonminimal functions) of the formulation switch is a double covering of said subset. It is precisely because of the structure of the group action of the formulation switch that the space of Palatini scalar-tensor theories is no richer than the space of ordinary metric scalar-theories. We emphasize that for wider classes of theories, or theories where the boundary terms are important, subtleties arise, and the two classes are not necessarily equivalent \cite{Dyer:2008hb}. For scalar-tensor theories with boundary at infinity, the presence of the non-minimal coupling does not widen the theory space.

\subsection{Metric-affine gravity and formulations of GR}

To the uninitiated, it may seem that a \emph{formulation} is nothing more than a representation of the same underlying theory, much like different conformal frames. If that were true, the literature would not abound with studies of various models in different formulations, all of which correspond to distinct physics. There is clearly more to the notion of a formulation, and it is important to examine it further before motivating an interpolation between different formulations.

When it comes to General Relativity, different formulations indeed are nothing more than equivalent representations of the same physics. This can be illustrated through the so-called geometric trinity of gravity \cite{BeltranJimenez:2019esp}, which contains GR, the teleparallel equivalent of General Relativity (TEGR) and the symmetric teleparallel equivalent of General Relativity (STEGR). Teleparallelism captures a wider array of theories than GR , but it is possible to project down to GR in the teleparallelism framework, which means that in the process, we arrive at different representations of GR.

To begin with, in GR, the degrees of freedom are encoded in the generalized curvature ${\cal R}^\alpha_{\ \beta\mu\nu}$. Under a metric-affine approach, the solutions to the Einstein--Hilbert action fix the connection to be Levi-Civita, and so the torsion and metricity do not have to be excluded by hand. All we have to do is write down the action in terms of the generalized Ricci scalar~$\mathcal{R}$:
\begin{align}\label{GRaction}
S_{\rm GR} =  \frac{1}{2}\int {\rm d}^4 x \sqrt{-g} \, \mathcal{R}.
\end{align}
This is the prototypical Einstein--Hilbert action, and a special case of \eqref{ehaction}, fully specifying ``pure'' GR.

Teleparallelism usually requires additional structure in terms of the frame bundle \cite{Aldrovandi:2013wha}, but it is possible to build up the so-called metric teleparallel theories by considering a general quadratic term constructed with the torsion $T^\rho_{\ \ \mu\nu}$:
\begin{align}
\mathcal{T}_{(c_1,c_2,c_3)} \equiv  -\frac{c_1}{4} T_{\rho\mu\nu} T^{\rho\mu\nu} - \frac{c_2}{2} T_{\rho\mu\nu} T^{\mu \rho\nu} + c_3 T^\rho T_{\rho}  \,,
\end{align}
where $T_\mu \equiv T^\rho_{\ \ \mu\rho}$.
Along with appropriate Lagrange multipliers, we may then use this to specify a three-parameter family of theories of gravity with vanishing non-metricity and curvature. TEGR is a special case in which $\mathcal{T} \equiv {\cal T}_{1,1,1}$. For this case, we can write
\begin{align}\label{relTR}
\mathcal{R} = R + \mathcal{T} + 2\nabla_\rho T^\rho.
\end{align}
As a result, through the flatness condition, we find that the following action recovers GR:
\begin{align}\label{TEGRaction}
S_{\rm TEGR} = -\frac{1}{2} \int {\rm d}^4 x \sqrt{-g} \, \mathcal{T},
\end{align}
since the surface term is as usual integrated away. Therefore, $S_{\rm TEGR}$ encodes the same dynamics as $S_{\rm GR}$,
%(note that it does not reduce to the same action since since the connection in $S_{\rm GR}$ is not assumed to be Levi--Civita)
with the difference that gravity is propagated by the torsion as opposed to curvature.

Similarly, we can build symmetric teleparallel theories by considering a general quadratic action constructed from the non-metricity $Q_{\rho\mu\nu}$:
\begin{align}
\mathcal{Q}_{c_1,c_2,c_3,c_4,c_5} \equiv \frac{c_1}{4} Q_{\rho\beta \gamma} Q^{\rho \beta \gamma} - \frac{c_2}{2} Q_{\rho \beta \gamma} Q^{\rho \alpha \gamma} - \frac{c_3}{4} Q_\rho Q^\rho + (c_4 - 1) \tilde{Q}_\rho \tilde{Q}^\rho + \frac{c_5}{2} Q_\rho \tilde{Q}^\rho \,,
\end{align}
for $Q_\rho = Q_{\rho \lambda}{}^{\lambda}$ and $\bar Q_\rho = Q^{\lambda}{}_{\lambda \rho}$. Again using appropriate Lagrange multipliers, we may use this to specify a five-parameter family of theories of gravity with vanishing torsion and curvature. STEGR is once again a special case in which $\mathcal{Q} \equiv \mathcal{Q}_{1,1,1,1,1}$. In this case, we can write
\begin{align}
\mathcal{R} = R + \mathcal{Q} + 2\nabla_\rho (Q^\rho - \bar Q^\rho).
\end{align}
Once again, with the flatness condition, we find that the following action recovers GR:
\begin{align}\label{STEGRaction}
S_{\rm STEGR} = -\frac{1}{2} \int {\rm d}^4 x \sqrt{-g} \, \mathcal{Q},
\end{align}
with the surface term integrated away. The argument is similar: $S_{\rm STEGR}$ encodes the same dynamics, only now gravity is propagated by the non-metricity.

\subsection{Models under different formulations}

The reason that TEGR and STEGR are known as formulations of GR is clear: each has a distinct action that encodes the same dynamics, something which is not the case for more general teleparallel theories. The same holds for the Palatini formulation of GR, which is distinct from the more general MAG theories. But given an action, what exactly does it mean to switch to a different formulation? Comparing the actions \eqref{GRaction}, \eqref{TEGRaction}, and \eqref{STEGRaction}, we see that formally, they only differ in the scalar term to be integrated. Therefore, it is established in the literature that changing the formulation of an action means to make a formal replacement of the associated scalar. For instance, starting from an action in the metric formulation defined in terms of~$R$, to switch to a different formulation is to make the replacement:
\begin{align}\label{formswitch}
R \to 
\begin{cases}
\hspace{0.8em} {\cal R} & \text{to Palatini} 
\\
{\cal -T} &  \text{to metric teleparallel} 
\\
{\cal -Q} &  \text{to symmetric teleparallel} 
\end{cases}
\end{align}
Such a replacement does not change the dynamics of a minimal action, even though the action does change. However, it is well known that this replacement does change the dynamics of a nonminimal action. In the teleparallel case, the dynamics of the resulting action cannot be encoded in a scalar-tensor theory, since the boundary terms now have a prefactor which not only means that they cannot be integrated away. Essentially, the action spaces of scalar-torsion models and nonminimal scalar-tensor models are not isomorphic, and the same goes for the action spaces of scalar-nonmetricity models.

It is a fortunate fact for the study of Palatini gravity that the action spaces of Palatini scalar-tensor gravity and metric scalar-tensor gravity indeed are, on the other hand, isomorphic: there are no additional dynamics or even model functions induced by relaxing the assumption that the connection takes on the Levi--Civita form. This leads to the well-known fact that after a switch from metric to Palatini (and vice versa), the dynamics of the resulting action can be encoded in a scalar-tensor action, even if this action will necessarily be situated at a different point (and in a different equivalence class) in the scalar-tensor action space.

It is very common to see in the literature statements to the effect of treating a particular \emph{theory} in some particular formulation, but with a greater understanding of the difference between actions and theories, we can see how these statements can be highly misleading. The choice of formulation is always made at the level of the \emph{action}, never at the level of a theory. We cannot present a theory in a different formulation without first selecting an action to represent it. It is meaningless to discuss the merits of a metric or Palatini or teleparallel formulation for a theory: we can only do so for an action, or, more precisely, a \emph{model}. Indeed, if by ``model'' we mean a particular set of functions that then go on to take on the role of the nonminimal coupling, the kinetic coupling, and the potential in some action, it makes perfect sense to talk about the theories induced by treating such a model in some particular formulation. This is in fact generally the sense in which the term ``model'' is used in practice (some examples are shown in Table~\ref{tab:models}): the models live ``above'' the action space, and are used to build up actions. In this way, we can specify (in analogy with the action space and the theory space) the \emph{model space} of scalar tensor theories $\mathbb{M}_{ST}$, its members are simply tuples of model functions, from which actions are constructed (diffrerently depending on the chosen formulation). 
\begin{table}
    \centering
    \begin{tabular}{c c c c}
        Model & $f(\phi)$ & $k(\phi)$ & $V(\phi)$ \\
        \hline
        Higgs inflation & $1+\xi \phi^2$ & 1 & $\frac{\lambda}{4}(\phi^2-v^2)$ \\
        Induced gravity inflation & $\xi\phi^2$ & 1 & $\frac{\lambda}{4}(\phi^2-v^2)$ \\
        nonminimal $\alpha$-attractors & $1 + \xi \phi^2$ & $\frac{1}{(\phi^2-6\alpha )^2}$ & $V(\phi)$ \\
         Brans--Dicke  & $\phi$ & $\frac{\omega_{\rm BD}}{\phi}$ & $V(\phi)$ \\
    \end{tabular}0
    \caption{\justifying Examples of single-field scalar-tensor models, each specified by its associated model functions. Once a model is chosen, then the choice of a formulation returns a specific action.}
    \label{tab:models}
\end{table}

Given the notions of models, actions, and theories, we can put them together by borrowing some elementary concepts from category theory. Focusing on metric scalar-tensor gravity to begin with, we specify the (concrete) category ${\cal C}_{\rm ST}^{\rm metric}$, whose objects are the model, actions, and theory spaces, and whose morphisms correspond to the mappings between the elements of the associated space (i.e. constructing the action and quotienting out the frame freedom). A commutative diagram is illuminating:
\begin{equation}
\begin{tikzcd}
\mathbb{M}^{\rm metric}_{\rm ST}
\arrow[r,"c_{\rm m}"] 
\arrow[rd,"b_{\rm m}"] 
\arrow[loop left, "{\rm id}^{\Omega,J^A{}_B}_\mathbb{M}"] 
& 
\mathbb{A}^{\rm metric}_{\rm ST}
\arrow[d,"t_{\rm m}"] 
\arrow[loop right, "{\rm id}^{\Omega,J^A{}_B}_\mathbb{A}"] 
\\
& 
\mathbb{T}^{\rm metric}_{\rm ST}
\arrow[loop right, "{\rm id}_\mathbb{T}"] 
\end{tikzcd}
\end{equation}
where we have the following morphisms:
\begin{itemize}
\item the ``action construction'' isomorphism $c_{\rm m}$ which maps from  model functions to  actions,
\item the ``theory'' epimorphism $t_{\rm m}$ that projects down to the quotient space by mapping actions to equivalence classes,
\item the composite ``model building'' epimorphism $b_{\rm m} = t_{\rm m} \circ c_{\rm m}$ that map a given set of model functions to a theory,
\item the identity morphism (and function since this is a concrete category) ${\rm id}_\mathbb{T}$ for the theory space, as well as the class of automorphisms ${\rm id}^{\Omega,J^A{}_B}_\mathbb{M}$ and ${\rm id}^{\Omega,J^A{}_B}_\mathbb{A}$ that relate map model functions and actions to their frame equivalents (with $\Omega = 1$ and $J^A{}_B = \delta^A_B$ specifying the identity function).
\end{itemize}

Analogously, we can define the category of Palatini gravity ${\cal C}_{\rm ST}^{\rm Palatini}$. We then note that ${\cal C}_{\rm ST}^{\rm metric}$ and ${\cal C}_{\rm ST}^{\rm Palatini}$ are equivalent categories. This is because of two reasons: first, we can define a functor $F$ that is both full and faithful (mapping the metric morphisms with $\rm m$ subscript to their Palatini equivalents with $\rm P$ subscript). Second, and crucially, $F$ is essentially surjective, i.e. it induces an isomorphism between the objects, i.e. the model, action, and theory spaces, since the theory space of metric and Palatini scalar-tensor gravity are isomorphic as we discussed earlier. Such an isomorphism, in practical terms, means that for a given model realized in metric, there exists a different model realized in Palatini such that these two models (with different functions) contain the exact same physical content. Indeed, it is the functor that induces such a bijective function between the metric and Palatini model spaces, and the explicit relation is derived in~\cite{Jarv:2020qqm}, (even if there is no closed-form algebraic relation in the general case). The categories for $F(\phi, R)$ theories are similarly equivalent, as long as we carefully count the degrees of freedom to map to the scalar-tensor category corresponding to the appropriate number of degrees of freedom  (since e.g. general $F(\phi,R)$ theories admit a biscalar equivalent).

The equivalence between categories is the formalization of the notion of a class of theories capturing the same physical content. We can take this further be seen by considering the categories of teleparallel scalar-tensor gravity as the categores of scalar-torsion and scalar-nonmetricity gravity.  While we can define a fully faithful functor between the previous categories and these ones, the scalar, the model, action, and theory spaces are not isomorphic as discussed in the previous section due to the existence of boundary terms. Therefore, the inequivalence of these categories is the formalization of the fact that that teleparallel formulations of scalar-tensor models (including scalar-torsion and scalar-nonmetricity theories) have a correspondingly larger theory space. In fact, scalar-tensor theories can be viewed as a subcategory of these wider theories. This can be seen once again through the lens of the number of invariant quantities needed to describe our models: scalar-tensor metric and Palatini gravity can be parametrized by the same set of such invariant quantities, whereas to describe scalar-torsion and scalar-nonmetricity gravity, this set must be augmented \cite{Hohmann:2018vle, Hohmann:2018dqh,Hohmann:2018ijr,Hohmann:2023olz}.

\begin{figure}[t]
    \centering
     \includegraphics[width=0.60\textwidth]{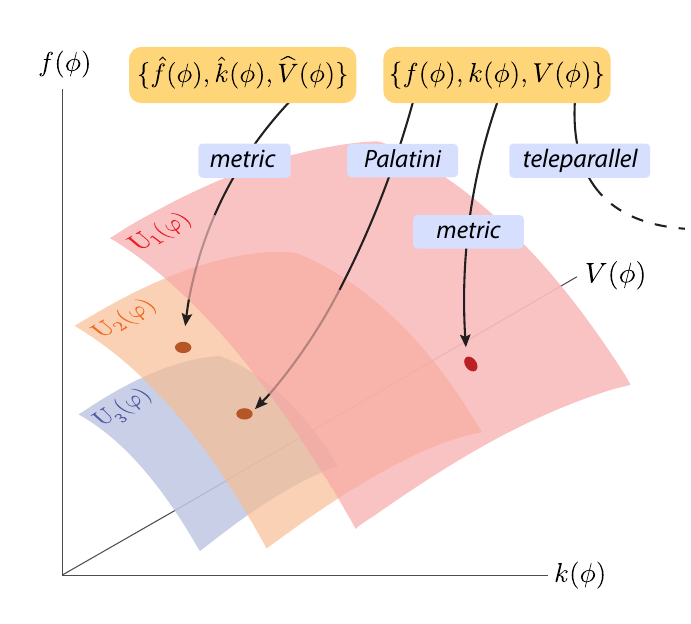} 
    \caption{\justifying 
    The space of single-field scalar-tensor metric actions, parametrized by the non-minimal coupling, kinetic coupling, and potential after both being written in terms of the usual Ricci scalar. The same set of model functions maps to different actions for the metric, Palatini, and teleparallel formulations (the latter of which belongs outside of the action space unless $f'(\phi) = 0$). Given a model interpreted as metric, there are multiple appropriately chosen models that point to actions corresponding to the same theory (i.e. the same equivalence class), and a single model that points to the exact same action.}
    \label{fig:models_equiv}
\end{figure}

In more practical terms, we can say this: it is an unfortunate choice of nomenclature that the procedure of writing down a specific action after selecting a particular model is associated with the selection of a formulation given that the resulting actions are not equivalent, but the terminology persists. Indeed, it has been pointed out that this is a naive approach to Palatini, as the so-called formulation swaps the theory for another \cite{Iglesias:2007nv}. Nonetheless, ``formulation'' in the literature overwhelmingly refers to the formal procedure of writing down a articular action  \emph{after} the choice of model parameters has been made, as illustrated in Fig.~\ref{fig:models_equiv}. Still, employing different formulations to motivate different theories of gravity has merit, and a continuous class of formulations can help further in this context as we shall see in the next section.   

\section{The quasi-Palatini formulation}
\label{quasi}

As we have discussed, metric and Palatini gravity are equivalent at the level of scalar-tensor theories. This however does not mean that the Palatini formualtion cannot lead to insights about gravity. Indeed, since different formulations of the same model can lead to distinct actions that hide different physics (i.e. are different theories), Palatini gravity holds ample potential for model-building, and multiple theories have been motivated by taking their metric counterparts and studying them in Palatini.

The notion of formulation as commonly presented in the literature is discrete. There have been successful attempts to bridge metric and Palatini gravity, notably in hybrid metric-Palatini gravity and $C$-theories, but such approaches are \emph{extensions} of a class of theories rather than distinct formulations of an underlying model.  Therefore, we propose a continuous family of formulations that interpolates between the metric and Palatini formulations, right after we briefly discuss hybrid theories and how they differ from our proposed approach.
 
\subsection{Hybrid metric-Palatini gravity}

Hybrid metric-Palatini gravity \cite{Harko:2011nh,Koivisto:2013kwa,Capozziello:2015lza} and the so-called $C$-theories \cite{Amendola:2010bk} are said to interpolate between metric and Palatini gravity. Given that both Einstein and Palatini gravity have advantages and disadvantages, combining the two makes it possible to have, in a sense, the best of both worlds (an example is the ability to have long-range forces in addition to being concordant with local gravity tests, which means that a screening mechanism is not required). Generalized hybrid Palatini theories \cite{Tamanini:2013ltp} are specified by an action of the form
\begin{align}
S = \frac{1}{2} \int \mathrm{d}^4 x \sqrt{-g} \, F(R, {\cal R}) + S_m
\end{align}
whereas $C$-theories are described similarly by by
\begin{align}
S = \frac{1}{2} \int \mathrm{d}^4 x \sqrt{-g} \, F({\cal R}_C)  + S_m
\end{align}
where ${\cal R}_C$ is defined with respect to a connection compatible with $C({\cal R})g_{\mu\nu}$. In both the above actions, the scalar-tensor representation is two-field. This all goes back to the important conceptual distinction between a formulation of a model and an extension of a theory: given a model function $F$ that can take $R$ or $\cal R$ as its argument, an interpolation between the metric and Palatini formulations should not necessitate the addition of new functions such as $F(R,{\cal R})$ or $C({\cal R})$. 

The situation for hybrid metric-Palatini theories, for which $F(R,{\cal R }) = M R + F({\cal R})$, is slightly different, as these do form an interpolation between two distinct classes of theories, and not between two formulatinons of the same theory. To see this, consider the following simple (metric) $F(R)$ action:
\begin{align}\label{FRmetric}
S_{\rm metric} &= \frac{1}{2} \int \mathrm{d}^4 x \sqrt{-g} \, F(R )  + S_m
\end{align}
After introducing an auxiliary field, we can write it in the following way:
\begin{align}
S_{\rm metric} &= \frac{1}{2} \int \mathrm{d}^4 x \sqrt{-g} \,\left[   F'(\Phi)R + F(\Phi)  -F'(\Phi) \Phi) \right] + S_m,
\end{align}
and through the redefinitions $\phi = F'(\Phi)$ and $2 V(\phi) = \Phi(\phi) F'(\Phi(\phi)) - F(\Phi(\phi))$:
\begin{align}\label{FRequivmetric}
S_{\rm metric} &= \frac{1}{2} \int \mathrm{d}^4 x \sqrt{-g} \,\left[   \phi R - 2V(\phi)\right] + S_m,
\end{align}
Similarly, the scalar-tensor realization of the equivalent Palatini action, where we replace $R\to {\cal R}$ in \eqref{FRmetric}, we find after replacing $R$ using \eqref{relationrR}:
\begin{align}\label{FRequivPalatini}
S_{\rm Palatini} &= \frac{1}{2} \int \mathrm{d}^4 x \sqrt{-g} \,\left[   \phi R + \frac{3}{2\phi} (\partial\phi)^2 - 2V(\phi)\right] + S_m.
\end{align}
Therefore, we may specify $F(R)$ and $F({\cal R})$ as Brans--Dicke type theories with $\omega_{\rm BD} = 0$ and $\omega_{\rm BD} = -3/2$, respectively. 

What we are looking for is a way to smoothly transition between the metric realization and the Palatini realization of a fixed model. In \cite{Capozziello:2015lza}, the following action was considered:
\begin{align}\label{hybridact}
S_\alpha = \frac{1}{2} \int {\rm d}^4 x \sqrt{-g} \, \left[ \alpha R+  F({\cal R})
\right] + S_m,
\end{align}
where $\alpha$ is a free parameter. For values of $\alpha\in [0,\infty)$, this action specifies a different point in action space, but the trajectory that it follows does not have the  $F({R})$ and $F({\cal R})$ actions at the endpoints. Indeed, the resulting action is
\begin{align}
S_\alpha = \frac{1}{2} \int {\rm d}^4 x \sqrt{-g} \, \left[ \alpha R + \phi {\cal R} - V(\phi)
\right] + S_m,
\end{align}
which, since the independent connection is compatible with $\phi g_{\mu\nu}$, can be written as the following equivalent scalar-tensor metric theory:
\begin{align}
S_\alpha =   \frac{1}{2} \int {\rm d}^4 x \sqrt{-g} \, \left[  (\alpha  + \phi) R + \frac{3}{2}\frac{(\partial  \phi)^2}{\phi} - 2V(\phi)
\right] + S_m.
\end{align}
Putting this action in the Brans--Dicke form, we find that the action is given by
\begin{align}\label{bdFR}
S_\alpha =   \frac{1}{2} \int {\rm d}^4 x \sqrt{-g} \, \left[ \phi R - \frac{\omega_{\rm BD}(\phi)}{\phi} (\partial  \phi)^2 - \frac{2\phi^2V(\phi)}{(\alpha+\phi)^2} 
\right] + S_m.
\end{align}
with
\begin{align}
\omega_{\rm BD}(\phi) = -\frac{3 \phi}{2 (\alpha+\phi )}.
\end{align}
As expected, when taking $\alpha \to 0$ (which means $\omega_{BD} = -3/2$), the Brans-Dicke equivalent of the hybrid action reduces to \eqref{FRequivPalatini}, which is to say that we are in the Palatini realization of the model $F({\cal R})$ (which is to be expected since the Einstein-Hilbert term in \eqref{hybridact} vanishes). However, for $\alpha\to \infty$ and $\omega_{BD}\to 0$, we find ourselves in GR, since the potential and kinetic terms are suppressed, which we could have already seen from the Hybrid action where this limit suppresses any Palatini contribution leaving us with GR with a non-dynamical effective Planck mass. After all, these hybrid theories can be parametrized by $X$, defined as
\begin{align}
X \equiv F(\mathcal{R})\mathcal{R}- 2 F(\mathcal{R})
\end{align}
which alternatively is given by the trace of the field equation:
\begin{align}
X = T + R.
\end{align}
Therefore, $X$ explicitly measures the deviation from GR and the trace equation $R=-T$.

\begin{figure}[t]
    \centering
     \includegraphics[width=0.60\textwidth]{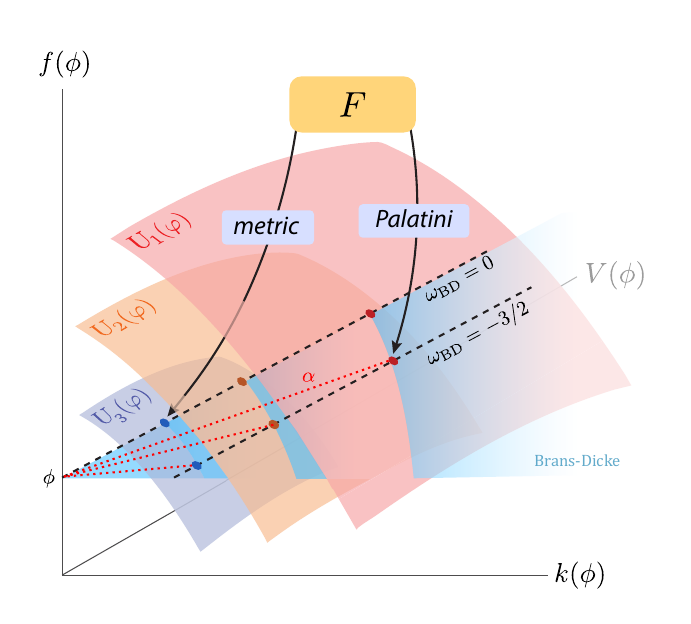} 
    \caption{\justifying 
    The classes of $F(R)$ and Palatini $F({\cal R})$ actions shown in the space of metric single-field scalar-tensor actions. Such actions live on the Brans-Dicke ``plane'' with $f(\phi) = \phi$. The dashed lines correspond to the constant values for the Brans--Dicke parameter ($0$ for metric and $-3/2$ for Palatini). Thus, $F(R)$ models, $F({\cal R})$ models, and any class of hybrid models with fixed $\alpha$ (schematically situated between the dashed lines) can be used to parametrize the space of theories (each point mapping to an equivalence class). As $\alpha$ varies, we move along the red dotted line, and the endpoints of the line do not correspond to the scalar-tensor equivalent of  $F(R)$ and $F({\cal R})$ for the \emph{same} function $F$, meaning that $\alpha$ is not a suitable candidate for a continuous deformation between formulations.
    }
    \label{fig:models_FR}
\end{figure}

What does all this mean for the hybrid action \eqref{hybridact} as a potential candidate for a way to make the notion of formulation continuous? To understand, it helps to look at the schematic representation of the situation in~Fig.~\ref{fig:models_FR}. For a given, fixed functional form of $F$, the metric and Palatini classes of actions define two separate lines in action space that do not meet. Given a fixed functional form for $F$, the different formulations of the same model point to different theories, just like in the scalar-tensor case, and the actions they point to have $\omega_{\rm BD} = 0$ for the metric $F(R)$ realization, and $\omega_{\rm BD} = -3/2$ for the Palatini $F(\mathcal{R})$ realization. Varying $\alpha$ and $F$ in the hybrid action \eqref{hybridact} does span the entire space between these classes of actions, but it does not interpolate between the metric and Palatini realization of the \emph{same} model with fixed~$F$.

All this should not be misconstrued as a claim that hybrid models such as \eqref{hybridact} are of limited use. Motivating theories with an additional parameter, even if said parameter does not broaden the space of theories, is a tried and true method of model building. Scalar-tensor theories themselves are an example: even if all the physics is encoded in the invariant potential $U(\phi)$, there is impetus from many sources (higher order corrections, the geometry of an underlying manifold, torsion/nonmetricity terms) to modify the nonminimal coupling or kinetic term, which changes the physics even though it does not lead to an entirely new class of theories. Hybrid metric-Palatini models represent a very useful avenue for model-building, but what they cannot do is interpolate between the metric and Palatini \emph{formulations} of the same \emph{model}.

\subsection{Interpolating between actions}

Looking again at the hybrid action \eqref{hybridact}, it is now evident that such a class of actions cannot interpolate between the two formulations of the same model since it only includes a $F({\cal R})$ term and not a $F(R)$ term. The parameter $\alpha$ does control the relative importance of the $R$ and $\cal R$ terms, which is what we want. Therefore, in order to achieve the same effect, and looking at the formulation switch procedure \eqref{formswitch}, we consider the following prescription:
\begin{align}\label{substR}
R \to R_{\delta_P} \equiv (1-\delta_P)R + \delta_P {\cal R},
\end{align}
where $\delta_P \in [0,1]$.  It is of course possible to affinely reparametrize $\mu \to \tilde \mu (\mu)$ with $\tilde \mu \in [0,\infty)$ in order to match the limits of $\alpha$ (such that now infinity corresponds to pure Palatini), but this parametrization has the advantage that it intuitively controls the relative ``Palatininess'' of the action, e.g. ``30 \% Palatini'' ($\delta_P = 0.3$) or ``$60\%$ metric'' ($\delta_P= 0.7$).

For scalar-tensor theories, this prescription leads to an action of the type
\begin{align}\label{FRquasi}
S_{\rm quasi} &= \int \mathrm{d}^4 x \sqrt{-g} \,  
\left[ 
\frac{(1-\delta_P) f(\phi) R}{2} 
+ \frac{\delta_P f(\phi) {\cal R}}{2} 
-\frac{k(\phi)}{2}(\partial\phi)^2 - V(\phi)
\right]
+ S_m.
\end{align}
The parameter $\delta_P$ has an intuitive interpretation of describing an action as e.g. ``30 \% Palatini'' ($\delta_P = 0.3$) or ``$60\%$ metric'' ($\delta_P= 0.4$).

We may now ask whether $\delta_P$ can be promoted to a function of the fields $\delta_P (\phi)$ or even a function of curvature $\Delta_P(R)$. We consider the following action
\begin{align}\label{FRquasi2}
S_{\rm quasi} &= \int \mathrm{d}^4 x \sqrt{-g} \,  
\left[ 
\frac{[1-\delta_P(\phi)] f(\phi) R}{2} 
+ \frac{\delta_P(\phi) f(\phi) {\cal R}}{2} 
-\frac{k(\phi)}{2}(\partial\phi)^2 - V(\phi)
\right]
+ S_m.
\end{align}
In this case, this is nothing more than an interpolation between the metric interpretation of a model $\{f_1,k, V\}$ and the Palatini interpretation of a model $\{f_2,k, V\}$ where
\begin{align}\label{FRquasi3}
f_1 (\phi) &= \frac{1-\Delta_P(\phi)}{1-\delta_P} f(\phi),
\\
f_2 (\phi) &= \frac{\Delta_P(\phi)}{\delta_P} f(\phi),
\end{align}
where $\delta_P$ still parametrizes the Palatininess of the model. To ensure that we are interpolating between the metric and Palatini realization of the same model is to set $f_1(\phi) = f_2(\phi)$, which returns $\Delta_P(\phi) = \delta_P$. A similar argument can be made for $\delta_P(R)$ in $F(R)$ models.

This argument demonstrates two points: first, if we wish to interpolate between the metric and Palatini formulations of the same model, we must encode the relative strength of Palatini as a pure number rather than a function. If we allow $\delta_P$ to have a dependence on $\phi$, that means we are interpolating between two actions that do not correspond to distinct formulations of the same model. However, this leads us to the second point: we can consider the even more general procedure of interpolating between \emph{any} two actions $S_1$ and $S_2$. The resulting action is nothing more than the weighted average:
\begin{align}\label{actinterp}
S_{\delta_S} = (1-\delta_S)S_1 + \delta_S S_2.
\end{align}
The interpolation defined in \eqref{actinterp} is as general as can be, given that $\delta_S$ cannot depend on the internal degrees of freedom of the theory. In this sense, it is an ``external'' interpolation whereas \eqref{substR} prescribes an ``internal'' interpolation. The intuitive meaning of $\delta_S$ is still the same: measuring the relative strength of the model. This prescription has the advantage can be defined at the level of the action, without any reference to the internal model functions or even the matter content of the theory. 

We note that the two actions $S_1$ and $S_2$ can truly be anything: they do not have to be related or share the same degrees of freedom, and the resulting interpolated action may have a more complicated internal structure than either $S_1$ or $S_2$. However, if the two actions are the different formulations of the same underlying model expressed in metric and in Palatini, respectively, then this interpolation defines a continuous class of formulations. We also note that the two prescriptions for interpolating return the same interpolated action for scalar-tensor actions where the minimal coupling takes on the form $f(\phi)R$. This is not the case for $F(R)$ actions, in which we find 
\begin{align} \label{FRint}
S^{\rm int}_{\rm quasi} &= \frac{1}{2} \int \mathrm{d}^4 x \sqrt{-g} \, F\big((1-\delta_P)R + \delta_P {\cal R} \big)  + S_m
\\ 
\label{FRext}
S^{\rm ext}_{\rm quasi} &= \frac{1}{2} \int \mathrm{d}^4 x \sqrt{-g} \, \Big[(1-\delta_P) F(R) + \delta_P F({\cal R} ) \Big]  + S_m.
\end{align}
Both of these actions can be viewed as the quasi-Palatini formulation of $F(R)$ gravity. We will examine both, along with the quasi-Palatini formulation of scalar-tensor gravity, in the upcoming subsections.

\subsection{Quasi-Palatini scalar-tensor gravity}

For scalar-tensor models, both the internal \eqref{substR} and external \eqref{actinterp} prescription return the same action, given in 
\eqref{FRquasi}. We now seek to write its equivalent metric form. Via a conformal transformation, we find that the resultant (single field) metric theory in the minimal frame becomes
\begin{align}
S_{\delta_P} = \int {\rm d}^4 x \sqrt{-g} \, \left[
\frac{R}{2} - \left(\frac{k(\phi)}{f(\phi)} + \frac{3(1-\delta_P)}{2}\frac{f'(\phi)^2}{f(\phi)^2}\right)
 (\partial\phi)^2 - U(\phi)\right],
\end{align}
where $U(\phi) = V(\phi)/f(\phi)^2$ as usual. This is a well-known form for the Einstein frame metric or Palatini action, depending on whether $\delta_P = 0$ or $\delta_P = 1$. The crucial difference is that $\delta_P$ now is no longer discrete, and can take any value in $[0,1]$. We further see that for $f'(\phi) = 0$, there is no distinction between metric and Palatini, as expected.

We assume that the kinetic prefactor is positive to avoid any ghosts. For $k>0$ and $\delta_P<1$, this is always the case: we could conceivably find ourselves in the ``anti-Palatini'' class of formulations where $\delta_P<0$. For $k>0$ and $\delta_P>1$ (the ``over-Palatini'' class of formulations), the existence of ghosts will depend on the form of $f(\phi)$ and $k(\phi)$. If the following bound is satisfied:
\begin{align}
\frac{k(\phi) f(\phi)}{f'(\phi)^2} > \frac{3(\delta_P-1)}{2}
\end{align}
for all $\phi$, then the theory is safe from ghosts even in the over-Palatini formulation. Still, we will restrict our attention to $0\le \delta_P \le 1$ in this work.

\begin{figure}[t]
    \centering
     \includegraphics[width=0.60\textwidth]{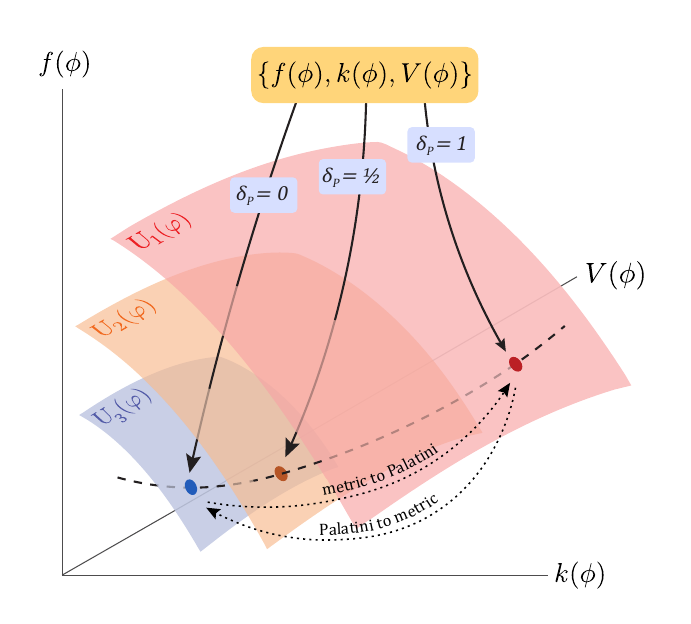}
    \caption{\justifying 
    The space of metric single-field actions with the points specified by a given model in different formulations. The value of $\delta_P$ now transforms continuously between actions and different sheets, each corresponding to a distinct invariant potential. Given an action identified as arising from a Palatini interpretation of some model, we can discretely switch to the action that corresponds to the same model interpreted in metric, which is the common discrete metric/Palatini duality of scalar-tensor gravity: the quasi-Palatini formulation provides a way to interpolate between these two actions. 
    }
    \label{fig:models_quasi2}
\end{figure}

From this point, we can transform to the Einstein frame by an appropriate redefinition of scalar:
\begin{align}
 \varphi(\phi)  = \int d\phi \sqrt{ \frac{k(\phi)}{f(\phi)} + \frac{3(1-\delta_P)}{2}\frac{f'(\phi)^2}{f(\phi)^2}}
\end{align}
Inverting this function defines a class of functions $\phi_{\delta_P}$, which can then be used to finally define a class of invariant potentials:
\begin{align}
U_{\delta_P} (\varphi) \equiv \frac{V(\phi_{\delta_P}(\phi))}{f(\phi_{\delta_P}(\phi))^2}.
\end{align}
As usual, $U_0(\varphi)$ is the ``pure metric'' potential and $U_1(\varphi)$ the ``pure Palatini'' invariant potential. These potentials encode different theories, in which $\delta_P$ appears as an additional parameter. This is illustrated in Fig.~\ref{fig:models_quasi2}.

\subsection{Quasi-Palatini $F(R)$}

First, we note that the theories given in \eqref{FRint} and \eqref{FRext}
are technically~$F(R,{\cal R)}$ theories. In the internal interpolation~\eqref{FRint}, we find that the Hessian of $F$ given by~$H_F$ vanishes:
\begin{align} 
H_F \equiv \frac{\partial^2 F}{\partial R^2} \frac{\partial^2 F}{\partial {\cal R}^2} - \left(\frac{\partial^2 F}{\partial R \partial{\cal R}}\right)^2,
\end{align}
which means that the parametrization in this case does not introduce an additional degree of freedom. Indeed, we can write the following action with the help of the usual auxiliary field:
\begin{align} 
S^{\rm int}_{\delta_P}= \frac{1}{2} \int {\rm d}^4 x \sqrt{-g} \left[ F(\Phi) + F'(\Phi)(R_{\delta_P}-\Phi)
\right]
\end{align}
Therefore, with the usual redefinitions $\phi = F'(\Phi)$ and $2 V(\phi) = \Phi(\phi) F'(\Phi(\phi)) - F(\Phi(\phi))$ we find
\begin{align} 
S^{\rm int}_{\delta_P} = \frac{1}{2} \int {\rm d}^4 x \sqrt{-g} \left[\phi R_{\delta_P} - 2 V(\phi) 
\right]
\end{align}
which expressed solely in terms of the metric through the relation between $R$ and $\cal R$ returns:
\begin{align}\label{FRinternalint}
S^{\rm int}_{\delta_P}= \frac{1}{2} \int {\rm d}^4 x \sqrt{-g} \left[\phi R + \frac{3 \delta_P}{2\phi}(\partial\phi)^2 - 2 V(\phi) 
\right].
\end{align} 
As we can see this model not only interpolates between the classes of metric and Palatini $F(R)$, but it interpolates precisely between the metric and Palatini realizations of a model with fixed $F$, since the potential does not change, unlike in \eqref{bdFR}. The Brans--Dicke coupling then varies from $-3/2$ for pure Palatini at $\delta_P = 1$ to $0$ for $\delta_P = 0$ as expected. This theory has exactly one dynamical scalar, just like the $F(R)$ metric theory. 

We now examine the external interpolation \eqref{FRext}. Here, the situation is slightly different: we find that the Hessian becomes
\begin{align} 
H_F = \delta_P(1-\delta_P)F''(R) F''({\cal R}).
\end{align}
This identically vanishes only if $F$ is linear, which corresponds to GR. Therefore, this formulation of the model always has an additional degree of freedom: it belongs to the class of generalized hybrid metric-Palatini gravity \cite{Tamanini:2013ltp}. In this case a single scalar field does not suffice, and we introduce two auxiliary fields $\Phi$ and $\Psi$ to write down the equivalent action, where $G(\Phi,\Psi) \equiv (1-\delta_P)F(\Phi)+\delta_P F(\Psi)$:
\begin{align} 
S^{\rm ext}_{\delta_P} = \frac{1}{2} \int {\rm d}^4 x \sqrt{-g} 
\left[
G(\Phi,\Psi) + \frac{\partial G}{\partial \Phi} (R-\Phi) + \frac{\partial G}{\partial \Psi} (R-\Psi).
\right]
\end{align}
%We need two auxiliary fields precisely because $H\ne0$, which implies that the system of equations for $\Phi$ and~$\Psi$ is not degenerate. 
Then, defining two new scalar fields $\phi \equiv \frac{\partial G}{\partial \Phi}$ and $\psi \equiv \frac{\partial G}{\partial \Psi}$, the action becomes
\begin{align} 
S^{\rm ext}_{\delta_P}  = \frac{1}{2} \int {\rm d}^4 x \sqrt{-g} 
\left[(1-\delta_P) \phi R + \delta_P \psi{\cal R} - 2V(\phi,\psi)
\right]
\end{align}
where the potential is simply
\begin{align}
V(\phi,\psi) =  (1-\delta_P) V(\phi) + \delta_P V(\psi)
%\phi \Phi(\phi) + \psi \Psi(\psi) - (1-\delta_P)F(\Phi(\phi)) -  \delta_P F(\Psi(\psi)) 
\end{align}
and
\begin{align}
2V(\phi) &= \phi \Phi(\phi) - F(\Phi(\phi))
\\
2V(\psi) &= \psi \Psi(\psi) - F(\Psi(\psi)).
\end{align}
The separability of the potential is guaranteed since the definitions of $\phi$ and $\psi$ give
\begin{align}
\phi &=  (1-\delta_P)F'(\Phi) 
\\
\psi &=  \delta_P F'(\Psi) 
\end{align}
and therefore inverting these equations guarantees $\Phi(\phi)$ and $\Psi(\psi)$ depend only on one of the fields.

Finally, using the relation between $\cal R$ and $R$, we find the action becomes
\begin{align}\label{FRRact}
S^{\rm ext}_{\delta_P}  = \frac{1}{2} \int {\rm d}^4 x \sqrt{-g} 
\left\{ \Big[ (1-\delta_P) \phi  + \delta_P \psi\Big] R - \frac{3\delta_P}{2\psi}(\partial\psi)^2 - 2 [ (1-\delta_P) V(\phi) + \delta_P V(\psi)] 
\right\}
\end{align}
This is a true two-field action with no redundant degrees of freedom, unlike the single-field action that arises as a result of the internal interpolation prescription: it has a two dynamical scalars, one more than the $F(R)$ metric theory. 

The two interpolations only match when $F(R,{\cal R})$ is linear in both arguments, i.e. $F$ is a linear function, which excludes them from the regular $F(R,{\cal R})$ treatment, since the Hessian vanishes. Such a linear function of course reduces to GR, and as such, the internal and external interpolations diverge for $F(R)$ theories. The same philosophy applies to $F(\phi,R)$ theories: we will find through analogous manipulations to the above that only when the function $F$ is linear in $F$ do the internal and external interpolation prescriptions match. For a general $F(\phi,R)$ theory, the internal interpolation leads to a theory of type $F(\phi,R+{\cal R})$, whereas the external interpolation to a theory of type $F(\phi,R)$+$F(\phi,{\cal R})$. The internally interpolated theory will have as many dynamical scalars as the regular $F(\phi,R)$ theory (just like with internally interpolated $F(R)$ theories), but the externally interpolated theory will have one more.

We finally note that both the internal and the external interpolation prescriptions can be equivalently applied by \emph{first} transforming the metric and Palatini theories to their Brans--Dicke equivalents \emph{separately} and \emph{then} interpolating between these (internally or externally, since it does not matter for $f(\phi)R$ scalar-tensor theories). There is one important distinction, however: in order to mimic the internal formulation interpolation of $F(R)$ theories, we must interpolate between the two Brans--Dicke theories with the same internal degrees of freedom, i.e. between $S_0[\phi,\partial\phi]$ and $S_1[\phi,\partial\phi]$ as given in \eqref{FRinternalint}:
\begin{align}
S^{\rm int}_{\delta_P}=  (1-\delta_P)S_{0}[\phi,\partial\phi] +   \delta_P S_{1}[\phi,\partial\phi] 
\end{align}
Doing so brings us simply to the general form $S_{\delta_P}[\phi,\partial\phi]$. On the other hand, if we wish to mimic the external formulation interpolation of $F(R)$ theories, we should interpolate between $S_{0}[\phi,\partial\phi]$ and $S_{1}[\psi,\partial\psi]$, which results in \eqref{FRRact} as desired thanks to the separability of the potential $V(\phi,\psi)$:
\begin{align}
S^{\rm ext}_{\delta_P} =  (1-\delta_P)S_{0}[\phi,\partial\phi] +   \delta_P S_{1}[\psi,\partial\psi] 
\end{align}
We therefore see that the internal interpolation corresponds to a projecting down of the line spanned by $\delta_P$, since if we identify $\psi = \phi$ in the externally interpolated action \eqref{FRRact}, we recover the internally interpolated action \eqref{FRinternalint}. 

\subsection{Quasi-Palatini discontinuity and conformal coupling}

Before we study a few models in the quasi-Palatini formulation, we note that the parametrization $\delta_P \in [0,1]$ may lead to discontinuities. This occurs, for instance, in $F(R)$ theories: it is well known that Palatini $F({\cal R})$ does not have a dynamical scalar degree of freedom, and as such, there is a rather stark jump between $\delta_P\ne1$ and $\delta_P = 1$. In this context, another parametrization of the interpolation may be more helpful, for instance one directly inspired by the hybrid models:
\begin{align}
S^{\rm int} &= \frac{1}{2} \int {\rm d}^4 x \sqrt{-g}\, \big[F(R) + \alpha F({\cal R}) \big],
\\
S^{\rm ext} &= \frac{1}{2} \int {\rm d}^4 x \sqrt{-g}\, F(R + \alpha {\cal R}).
\end{align}
With these interpolations, the discontinuity is shifted to $\alpha \to \infty$. However, since there is no such discontinuity for $f(\phi)R$ actions, which do have a Palatini interpretation with a dynamical scalar field, the $\delta_P$ parametrization is more convenient. 

Another instance in which the discontinuity is apparent is when considering the conformal couping. A field conformally coupled to the metric Ricci scalar $R$, has an action in $D$ dimensions given by
\begin{align}
S =   \frac{1}{2}\int \mathrm{d}^D x \, \sqrt{-g} \left\{ \partial_\mu \phi \partial^\mu \phi - \xi \phi^2 [(1-\delta_P) R + \delta_P {\cal R}]\right\}\, 
\end{align}
for a particular value of $\xi$. We now consider a Weyl transformation, assigning a conformal weight of $\Delta$ to the field. The infinitesimal transformations are given by
\begin{align}
\delta g_{\mu\nu} &\approx  2\omega g_{\mu\nu}
\\
\delta \phi &\approx  \Delta \omega \phi,
\\
\delta{R}_{\delta_P} &\approx   -2\omega R - \Delta \gamma_\mu \nabla^2 \omega,
\end{align}
where the prefactor $\gamma_{\delta_P}$ for the transformation of $R_{\delta_P}$ is given by
\begin{align}
\gamma_\mu \equiv \frac{4 (1-\delta_P)(D-1)}{(D-2)}.
\end{align}
Further using $\delta g^{\mu\nu} = -2\omega g^{\mu\nu}$ and $\delta \sqrt{-g} = D \omega \sqrt{-g}$, we find:
\begin{align}\label{confvar2}
 2\delta S =  \int \mathrm{d}^D x \,  \omega \sqrt{-g}\left[
\left(D-2 +2 \Delta \xi  \gamma_{\delta_P} \right)    (\nabla  \phi )^2
+2 \Delta\left(   \xi  \gamma_{\delta_P}   -   1 \right) \phi\nabla^2\phi 
-\xi \left( 2 \Delta - 2 +  D  \right)\phi^2 R_{\delta_P}  
 . \right] 
\end{align}
The terms vanish identically if $\Delta = 1-D/2 $, as usual, and
\begin{align}
  \xi_{\delta_P} &= \gamma_{\delta_P}^{-1} = \frac{D-2}{4(1-\delta_P ) (D-1)}.
\end{align}
We of course find as usual that for $\delta_P = 0$ (the fully metric theory), the conformal coupling is given by the usual expression, taking on the usual critical value $\xi_c = -1/6$ for $D=4$ from the general case $\xi_c = -\frac{1}{6(1-\delta_P)}$. Furthermore, the discontinuity between Palatini and metric is once again made explicit as $\delta_P = 1$ makes the coupling blow up, reflecting the fact that $\gamma_1$ = 0, which means that \eqref{confvar2} cannot vanish identically and a $\xi\phi^2 \mathcal{R}$ coupling cannot make the action conformally invariant in Palatini. 

\subsection{Beyond scalar-tensor}

The idea of interpolating between formulations does not have to be restricted to scalar-tensor theories. For instance, the interpolation between the regular metric interpretation of a model specified by $f(\phi)$, $k(\phi)$ and $V(\phi)$ and the symmetric teleparallel interpretation leads to a scalar-torsion action:
\begin{align}
S = \frac{1}{2} \int d^4 \sqrt{-g} \, 
\left\{f(\phi) \big[ (1-\delta_{\cal T})R - \delta_{\cal T} {\cal T}\big] - k(\phi)(\partial\phi)^2 - 2 V(\phi) \right\} ,
\end{align}
where the term $ (1-\delta_T)R - \delta_T T$ replaces the usual Ricci scalar, and $\delta_T$ now represents the ``torsion-ness'' of the action. Once again, for $f(\phi)$ constant, there is no distinction between the formulations. However, in general, the presence of the nonminimal coupling does more than just point to a different scalar-tensor action. Unlike the quasi-Palatini formulation, this ``quasi-torsion'' formulation takes us out of the scalar-tensor action space as specified by \eqref{scaltensact}, since the $2f(\phi) \nabla_\rho T^\rho $ term from \eqref{relTR} can no longer be integrated away. This formulation takes us to the more general class of scalar-torsion theories, specified by the scalar-torsion action
\begin{align}
S = \frac{1}{2} \int d^4 \sqrt{-g} \, 
\left\{f(\phi) \big[ (1-\delta_T)R - \delta_{\cal T} {\cal T}\big] - k(\phi)(\partial\phi)^2 - 2 V(\phi) \right\}.
\end{align}
Such a theory is then equivalent to the scalar-torsion action
\begin{align}
S = \frac{1}{2} \int d^4 \sqrt{-g} \, 
\left\{-f(\phi) {\cal T}  - k(\phi)(\partial\phi)^2 + 2(1- \delta_{\rm T}) f'(\phi)   T^\mu  (\partial_\mu \phi)- 2 V(\phi) \right\},
\end{align}
which corresponds to a subclass of scalar-torsion theory \cite{Hohmann:2018ijr, Hohmann:2023olz} specified by three rather than the general four model functions, since the torsion prefactor $C(\phi)$ is induced by the derivative of the nonminimal coupling $f(\phi)$. For the purely torsion formulation, this term vanishes, although an equivalent purely scalar-tensor description does not exist. Thus, for a scalar-tensor model defined by three functions, its quasi-torsion formulation does not live in the scalar-tensor action space, unless $f(\phi)$ is a constant. A similar construction can be made for the ``quasi-metricity'' formulation of a model, in which case we will find that again, there is no equivalent scalar-tensor description. 

In general, there is no need to restrict ourselves to weighting just a pair of formulations. For instance, given a usual scalar-tensor model, we can consider its quasi-teleperallel formulation, given by the substitution
\begin{align}
R \to \delta_{\mathcal{R}} \mathcal{R} - \delta_{\cal T} {\cal T} + \delta_{Q} {\cal Q}, 
\end{align}
where $\delta_{\cal R} + \delta_{\cal T} + \delta_{\cal Q} = 1$. Again, given a minimal starting model with $f(\phi)$ constant, this substitution does not give us any physics beyond GR. However, in the presence of a nontrivial nonminimal coupling, new degrees of freedom are introduced, representing a rich potential avenue for model-building.

\section{Quasi-Palatini scalar-tensor inflation}
\label{infl}

We will now examine some more practical applications of the quasi-Palatini gravity formulation. We will first examine the cosmological equations of motion, and then we will examine a few models of inflation and how these fare under the quasi-Palatini class of formulations. We note that in this section, we will focus on the scalar-tensor formulation, and so $F(R)$ models are implicitly assumed to be studied in the internal quasi-Palatini interpolation, in ordet to keep the number of dynamical degrees of freedom minimal.

We consider the model specified by $f(\phi)$, $k(\phi)$, and $V(\phi)$, as well as an associated matter action $S_m$ in the quasi-Palatini formulation. The action is described by \eqref{FRquasi}. By eliminating $\cal R$, we can write the Einstein equations in terms of metric-defined quantities:
\begin{align}
f(\phi) G_{\mu\nu} = T^{\rm eff}_{\mu\nu}
\end{align}
where the effective energy-momentum tensor can be broken up into three components, the matter component (originating as usual from $S_m$), the scalar-tensor component (originating from the effective fluid description of the scalar field and includes term common to both metric and Palatini), and the metric scalar-tensor component, which involves derivatives of the nonminimal term and appears \emph{only} in the metric formulation:
\begin{align}
T^{\rm eff}_{\mu\nu} \equiv   T^{\rm m}_{\mu\nu} + T^{\rm ST}_{\mu\nu} + (1-\delta_P)T^{\rm ST,m}_{\mu\nu},
\end{align}
where
\begin{align}
T^{\rm m}_{\mu\nu}  &\equiv -\frac{2}{\sqrt{-g}} \frac{\delta S_m}{\delta g^{\mu\nu}}
\\
T^{\rm ST}_{\mu\nu} &\equiv k(\phi) \nabla_\mu \phi \nabla_\mu \phi - \frac{g_{\mu\nu} k(\phi)}{2}(\nabla\phi)^2+ g_{\mu\nu}V(\phi)
\\
T^{\rm ST,m}_{\mu\nu} &\equiv  f''(\phi) \nabla_\mu\phi \nabla_\nu\phi -g_{\mu\nu} f''(\phi) (\nabla\phi)^2 +f'(\phi) \nabla_\mu \nabla_\nu \phi   - g_{\mu\nu} f'(\phi) \nabla^2 \phi.
\end{align}
The scalar equation of motion is given by
\begin{align}
k(\phi) \nabla^2\phi + \frac{k'(\phi)}{2} (\nabla\phi)^2  + V'(\phi)+ \frac{(1-\delta_P)f'(\phi)}{2}R  = 0,
\end{align}
where we remind that we have written everything in terms of the usual metric Ricci scalar. 

In practice, it is easier to work in the Einstein frame. After a conformal transformation, we assume a FLRW background $ds^2 = -dt^2 + a(t)^2 \delta_{ij}dx^i dx^j$, and we can write the resulting cosmological equations
\begin{align}
3 H^2 &=  \frac{G_P(\phi)}{2}     \dot\phi^2 + \frac{V(\phi)}{f(\phi)} + \frac{\rho_m}{f(\phi)} ,
\\
\dot H &= -  \frac{G_P(\phi)}{2}  \dot\phi^2 - \frac{\rho_m +p_m}{2f(\phi)}
\end{align}
where as usual $H\equiv \dot a/a$, with the scale factor $a$ being defined in the Einstein frame. We also can define $\rho_m$ and $p_m$ through from the fluid description of the matter sector, and $G_P(\phi)$ is just the single-field field space metric given in \eqref{fieldmetric}. Finally, the continuity equation along with $R = 12H^2 + 6 H^2$ leads to
\begin{align}
\ddot\phi + \frac{G_P'(\phi)}{2G_P(\phi)}\dot\phi+ 3H \dot\phi + \frac{U'(\phi)}{G_P} +  \frac{(1-\delta_P)}{2}\frac{f'(\phi)}{f(\phi) G_P(\phi)} ( \rho_m - 3p_m) = 0,
\end{align}
where the second term comes from the Levi-Civita connection for $G_{AB}$, although in the single-field case, it can be fully eliminated by an appropriate field redefinition, since the field-space manifold is trivial, even if such a redefinition may be impractical. We note that the last term vanishes in the case of the pure Palatini formulation, which means that the Klein--Gordon equation is fully decoupled from the matter continuity equation $\dot\rho_m + 3H(\rho_m + p_m) = 0$. 

We will now proceed to examine some concrete examples of the quasi-Palatini formulation by interpolating between metric and Palatini for some well-known models of inflation. We note that hybrid metric-Palatini gravity has been considered in the context of inflation \cite{He:2022xef}, but such approaches do not interpolate between different formulations of the same model. We focus on inflation precisely because we are able to draw from the powerful machinery available to link theory and predictions, therefore clearly demonstrating the effects of interpolating between different models.

\subsection{Higgs inflation}

As shown in Table~\ref{tab:models}, Higgs inflation is specified by a second-order nonminimal coupling and a symmetry-breaking quartic potential \cite{Bezrukov:2007ep}. The VEV of the potential is very small, and so we can write the action as 
\begin{align}
S = \int {\rm d}^4 x \sqrt{-g} \left\{\frac{1+\xi\phi^2}{2}\big[(1-\delta_P)R + \delta_P {\cal R}\big]+ \frac{(\partial\phi)^2}{2}- \frac{\lambda}{4}\phi^4\right\}
\end{align}
where we assume that during inflation the matter sector has been diluted by the scalar field. 

The most straightforward way to proceed is to first eliminate the nonminimal coupling via a conformal transformation, and then canonicalize the field from $\phi$ to $\varphi$ in order to define the invariant potential $U(\varphi)$ in terms of it. The canonicalization equation in this case is
\begin{align}
\frac{d\varphi}{d\phi} = \sqrt{\frac{1}{1+\xi  \phi ^2}+\frac{6 \delta_m  \xi ^2 \phi ^2}{\left(1+\xi  \phi ^2 \right)^2}}.
\end{align}
where for convenience, we have defined $\delta_m = 1-\delta_P$, which, dual to $\delta_P$, encodes the ``metricness'' of the action. It is thankfully possible to solve this equation in closed form:
\begin{align}\label{phiinv}
\varphi  = \frac{\sqrt{1+6  \delta_m \xi } \, \sinh ^{-1}\left(\sqrt{\xi } \phi  \sqrt{1+6 \delta_m  \xi }\right)}{\sqrt{\xi }}-\sqrt{6  \delta_m}   \tanh ^{-1}\left(\frac{\sqrt{6  \delta_m} \, \xi  \phi }{\sqrt{1+\xi  \phi ^2+ 6   \delta_m  \xi ^2 \phi ^2 }}\right).
\end{align}
Inverting this equation defines a function $\phi_{\delta_P} (\varphi)$, which in turn defines the invariant potential
\begin{align}
U_{\delta_m}(\varphi) = \frac{\lambda}{4} \big[\phi_{\delta_m}(\varphi)\big]^4.
\end{align}
However, it is unfortunately not possible to invert \eqref{phiinv} in closed form. Nonetheless, we may achieve some intuition by considering $\phi\gg1$, which corresponds to the plateau where inflation takes place. In this case, we find
\begin{align}
\varphi_{\delta_m}(\phi) =\frac{1}{\sqrt{\xi } \sqrt{1+6 \delta_m \xi}} \ \sinh \left[\frac{\sqrt{\xi } (\phi+\gamma   )}{\sqrt{1+6 \delta_m  \xi  }}\right],
\end{align}
where
\begin{align}
\gamma \equiv \sqrt{6 \delta_m}   \tanh ^{-1}\left(  \xi  \sqrt{\frac{6\delta_m }{  (1+6 \delta_m   \xi  )\xi}}\right).
\end{align}
Again, we note the discontinuities present when the conformal coupling value $\xi = -1/(6\delta_m)$ is achieved. Further expanding for $\phi\gg1$, we find that $\varphi_{\delta_m}(\phi)$ becomes
\begin{align}
\varphi_{\delta_m}(\phi) \approx \frac{1}{2 \sqrt{\xi } \sqrt{1+6 \delta_m \xi}}
\exp\left[ {\frac{\sqrt{\xi } (\phi+\gamma )}{\sqrt{1+6 \delta_m  \xi }}}\right],
\end{align}
which finally returns the large-field approximation for the Einstein frame potential $U(\phi) = V(\phi)/f(\phi)^2$ in terms of $\varphi$
\begin{align}
U_{\delta_m}(\varphi)=\frac{\lambda  e^{\frac{4 \sqrt{\xi } (\gamma +\phi )}{\sqrt{1+6 \delta_m  \xi }}}}{4 \xi ^2 \left(e^{\frac{2 \sqrt{\xi } (\gamma +\phi )}{\sqrt{1+6 \delta_m  \xi }}}+24 \delta_m  \xi +4\right)^2}.
\end{align}
We note that this defines a class of Einstein-frame potentials that interpolates between the metric and Palatini potentials for the fixed functional form of the non-minimal coupling.

We can now proceed to determining the observables. As usual, the observables at horizon exit are determined by the Hubble slow-roll parameters
\begin{align}
\epsilon_H \equiv -\frac{\dot H}{H^2},
\qquad
\eta_H \equiv \frac{\dot{\epsilon_H}}{H\epsilon_H},
\end{align}
along with the definition of the number of e-foldings $dN = H dt$.
The second-order slow-roll parameter is defined in various ways throughout the literature, but the recursive definition (in terms of $\epsilon_H$ lends itself to a straightforward definiton of the higher order slow-roll parameters. We focus on the three ``big'' observables: the spectral index $n_s$, the tensor-to-scalar ration $r$, as well as the strength of the scalar spectrum $A_s$. These are given as follows:
\begin{align}
n_s &= 1 - 2\epsilon_H + \eta_H,
\\
r &= 16\epsilon_H,
\\
A_s &= \frac{32}{\pi^2} \frac{H^2}{r}.
\end{align}
It is possible to define a set of potential slow-roll parameters $\epsilon$ and $\eta$ that reduce exactly to the Hubble slow-roll parameters in the slow-roll limit. The deviation from slow-roll that occurs in the phase space of inflation affects the observables chiefly in two ways: first, by shifting the end of inflation, which is precisely defined by $\epsilon_H = 1$ and approximately by $\epsilon =1$, and secondly by introducing higher order terms in the relation between the potential and the Hubble slow-roll parameters $\epsilon = \epsilon_H + {\cal O}(\epsilon_H^2,\epsilon_H\eta_H,\eta_H^2)$. These deviations are expected to become more important as cosmological precision increases in differentiating between models, but currently they are still subleading.   

The potential counterparts to the Hubble slow-roll parameters are given by
\begin{align}
\epsilon  \equiv \frac{1}{2}\left[\frac{U'(\varphi)}{U(\varphi)}\right]^2,
\qquad
\eta \equiv \frac{\epsilon'(\phi) }{\epsilon(\phi)}\frac{U'(\phi) }{U(\phi)}.
\end{align}
and the observables can be expressed in terms of $\varphi$ as
\begin{align}
n_s &= 1 - 2\epsilon_U + \eta_U,
\\
r &= 16\epsilon_U,
\\
A_s &= \frac{2}{3} \frac{U}{\pi^2\epsilon}
\end{align}
We may also approximate the number of e-foldings via
\begin{align}
\frac{dN}{d\phi} = \frac{U(\phi)}{U'(\phi)},
\end{align}
which allows us to express the observables in terms of $N$ (although in practice it is not possible to invert the expression for $N(\varphi)$, necessitating the use of approximations).

\begin{figure}[t]
    \centering
    \includegraphics[width=0.40\textwidth]{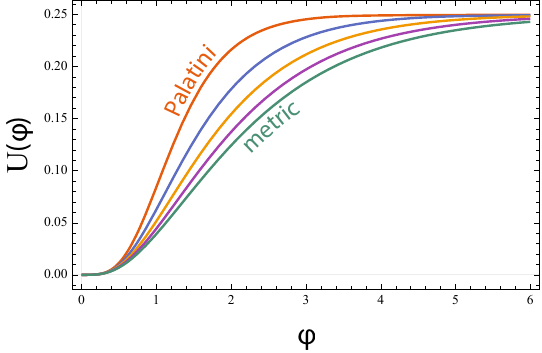} \qquad
     \includegraphics[width=0.40\textwidth]{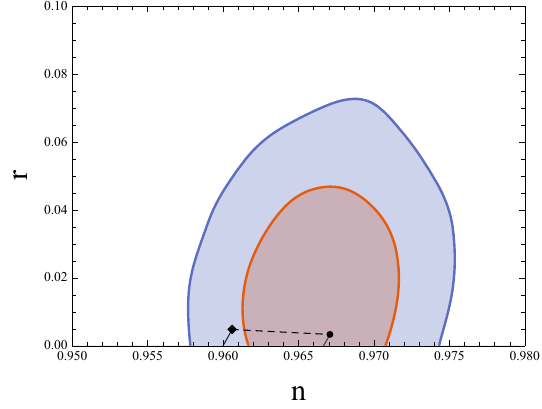}
    \caption{\justifying
    {\bf Left:} normalized invariant potentials for quasi-Palatini Higgs inflation in intervals of $0.25$ for $\delta_m$ plotted against the Planck contours at $68\%$ and $95\%$ confidence. The Palatini potential is the steepest, whereas the metric is the shallowest, although they all have the same general sigmoid shape. {\bf Right:} the predictions of quasi-Palatini Higgs inflation. The dashed line corresponds to pure metric inflation (the diamond and circle correspond to $N_* = 50$ and $N_* = 60$. respectively). }
    \label{fig:higgs}
\end{figure}

For a small field value at the end of inflation, we can invert $N(\varphi)$, from which we have
\begin{align}
\phi = 
\frac{1}{2} \left[\frac{\sqrt{1+6 \delta_m \xi  } \, \ln (32 \xi N)}{\sqrt{\xi }}-2 \gamma \right].
\end{align}
This finally returns the following expressions for the spectral index and tensor-to-scalar ratio:
\begin{align}
n_s &= \frac{128 \xi  (1+ 6 \delta_m \xi )}{1+6 \delta_m \xi +8 \xi  N^2} = 1- \frac{2}{N} + \frac{1+6 \delta_m \xi}{4 \xi  N^2} +  {\cal O}(N^{-3}),
\\
r &= \frac{1+2 \xi  (3 \delta_m-8+4 N)}{1+6 \delta_m \xi +8 \xi N} = \frac{2 (1+6 \delta_m \xi )}{\xi  N^2} +  {\cal O}(N^{-3}).
\end{align}
We note that $\delta_m$ can indeed be used to interpolate between the usual predictions of metric Higgs inflation and Palatini inflation. However, we remind that $\xi$ depends on $\delta_m$ through the normalization of the spectrum. Writing $A_s$ in terms of $N$, we find:
\begin{align}\label{spectnorm}
A_s = \frac{\lambda  N^2}{72 \pi ^2 \delta_m  \xi ^2+12 \pi ^2 \xi }.
\end{align}
The most recent observational constraints \cite{Planck:2018vyg} give $A^*_s = (2.1 \pm 0.0589)\times10^{-9}$ in Planck units. Taking $N_*$ to vary between 50 and 60 e-foldings, we again find the usual values for $\xi$, ${\cal O}(10^4)$ in Higgs inflation and ${\cal O}(10^9)$ in Palatini inflation. Solving for $\xi$ in terms of $\delta_m$, we can finally show the contour plot, with one crucial difference: the value of $\xi$ can now be used to probe the relative Palatininess of the model, with the tensor-to-scalar ratio as usual driven down in predominantly Palatini Higgs due to the large value of $\xi$. We illustrate the predictions in Fig.~\ref{fig:higgs}, where we note that mixing Palatini with Higgs deforms the invariant potential, bringing down the tensor-to-scalar ratio prediction.

\subsection{Induced gravity inflation}

\begin{figure}[t]
    {
    \centering
    \includegraphics[width=0.40\textwidth]{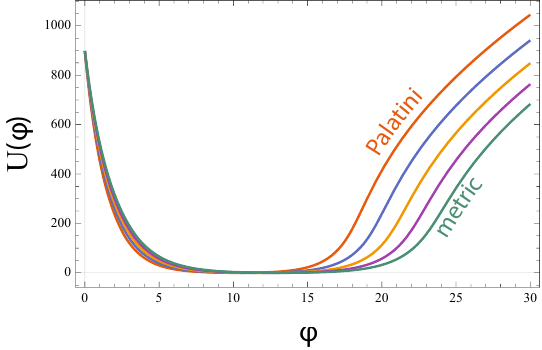} \qquad
     \includegraphics[width=0.40\textwidth]{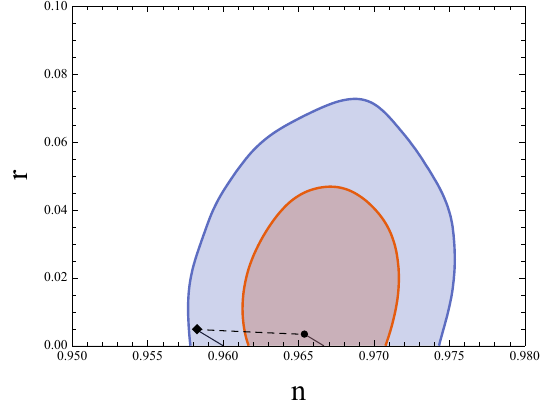}
    }
    \caption{\justifying 
    {\bf Left:} normalized invariant potentials for quasi-Palatini induced gravity inflation in intervals of $0.25$ for $\delta_m$. For small field values, the potentials nearly overlap, though they do give rise to differentiated predictions. {\bf Right:} the predictions of quasi-Palatini induced gravity inflation plotted against the Planck contours at $68\%$ and $95\%$ confidence. The dashed line corresponds to pure metric inflation (the diamond and circle correspond to $N_* = 50$ and $N_* = 60$. respectively). }
    \label{fig:induced}
\end{figure}

For induced inflation \cite{Fakir:1990eg}, much of the procedure is similar. We start with the action, which interpolated between metric and Palatini gives us 
\begin{align}
S = \int {\rm d}^4 x \sqrt{-g} \left\{\frac{ \xi\phi^2}{2}\big[(1-\delta_P)R + \delta_P {\cal R}\big]+ \frac{(\partial\phi)^2}{2}- \frac{\lambda}{4}(\phi^2 - v)^2\right\}.
\end{align}
Unlike in Higgs inflation, we cannot ignore the VEV, as settling down in it is what sets the Planck mass. Indeed, we find that $v = 1/\sqrt{\xi}$. Then, through a process analogous to that of the previous section, we determine the
\begin{align}
U_{\delta_m}(\varphi)=\frac{1}{\xi ^4} e^{-\dfrac{4 \sqrt{\xi } \phi }{\sqrt{1+6 \delta_m \xi }}} \left(\xi  e^{\dfrac{2 \sqrt{\xi } \phi }{\sqrt{1+6 \delta_m \xi }}}-1\right)^2.
\end{align}
In this case, we consider the small field expansion, for which
\begin{align}
\phi(N) = \frac{\sqrt{1+6 \delta_m \xi  } }{2 \sqrt{\xi }} \log \left(\frac{8 N}{1+6 \delta_m \xi}\right).
\end{align}
With this, we can calculate the observable quantities in terms of $N$ as
\begin{align}
n_s &= 1 -\frac{2}{N}
-\frac{3 (1+6 \delta_m \xi )}{4 \xi  N^2} + {\cal O}(N^{-3}), 
\\
r &= \frac{2 (1+6 \delta_m \xi )}{\xi  N^2} + {\cal O}(N^{-3})
\end{align}
where once again $\xi$ must be determined from the normalization of the spectrum:
\begin{align}
A_s = \frac{(1+6 \delta_m \xi -8 \xi N )^4}{12288 \pi ^2 \xi ^5 N^2 (1 + 6 \delta_m \xi )}.
\end{align}
Using the observed value for $A_s$, we can plot the predictions on Fig.~\ref{fig:induced}: we note that the gradient of the $\delta_m$ line is opposite to that of Higgs inflation. Once again, the variation in the potential is enough to lead to differentiated predictions at second order.

\subsection{Starobinsky inflation}

We finally consider the Starobinsky model, described by $F(R) = R + \beta R^2$. Putting in the form given by \eqref{FRequivPalatini}, we find 
\begin{align} 
S_{\rm Palatini} &= \frac{1}{2} \int \mathrm{d}^4 x \sqrt{-g} \,\left[   \phi R + \frac{3\delta_P}{2\phi} (\partial\phi)^2 - \frac{(\phi-1)^2}{4\beta}\right] + S_m.
\end{align}
Note that this is distinct from the external interpolation, which features two degrees of freedom:
\begin{align} 
S_{\rm Palatini} &= \frac{1}{2} \int \mathrm{d}^4 x \sqrt{-g} \,\left[   (\delta_m\phi + \delta_P\psi)R + \frac{3\delta_P}{2\phi} (\partial\phi)^2 - \frac{\delta_m}{4\beta}(\phi-1)^2 + \frac{\delta_P}{4\beta}(\psi-1)^2\right].
\end{align}
Still, as we mentioned at the start of this section, we will focus on the models with one degree of freedom. By canonicalizing the field as usual, we find
\begin{align}
\phi = \exp  \sqrt{\frac{2}{3\delta_m}} \ \varphi,
\end{align}
from which we move on to writing the invariant potential as
\begin{align}
U(\varphi) = \frac{1}{2 \beta } \left(1-e^{-\sqrt{\frac{2}{3 \delta_m}} \varphi } \right)^2
\end{align}
From this potential, we can see the role that the quasi-Palatini interpretation plays: it leads to a deviation from the regular Starobinsky model by modifying the residue of the pole found in the action.

Finding the number of e-foldings for large field values and inverting, we find
\begin{align}
n_s &= \frac{3 \delta_m (3 \delta_m-8)+16N^2-8 (3 \delta_m+4)N}{(4N-3 \delta_m)^2}
\\
r&= \frac{192 \delta_m}{(4 N-3 \delta_m)^2},
\end{align}
or, to second order:
\begin{align}
n_s &= 1-\frac{2}{N} -\frac{9 \delta_m}{2 N^2}
\\
r&= \frac{12 \delta_m}{N^2}.
\end{align}
These match the $\alpha$-attractor predictions \cite{Galante:2014ifa}, with the added benefit that the predictions are robust with respect to $\beta$, which does not enter into the predictions at any order, since it only appears as a normalization factor. Therefore, by ``doping'' the model with Palatini (without making it solely Palatini otherwise there would be no dynamics), we can arbitrarily suppress the tensor-to-scalar ratio.

\section{Conclusion}
\label{concl}

For all intents and purposes, the selection of a model begins by specifying a set of functions, whose form may be motivated by higher-order terms, radiative corrections, or inherited by specific particle physics models, among many other possible motivations. This corresponds to an overdetermination of the theory space: it is possible to write multiple distinct actions that nonetheless contain the same physics. The geometric trinity of gravity is a clear example of the overabundance of distinct descriptions of the same physics. Different actions constructed out of fundamentally distinct objects (curvature, torsion, non-metricity) form a wider class of theories than general relativity, but upon a careful choice of parameters, they are equivalent to it. This equivalence is the impetus for the idea of a formulation: a different realization of a model with fixed functional parameters that is merely a reparametrization of the same physics in the ``usual'' case (minimal theories), but which represents a tangible physical change in every other case. Therefore, it is an unspoken convention that models exist in a sense independent of their formulation, and that for all models save for the minimal ones, different formulations will lead to different physics (despite what the term might imply). 

For scalar-tensor gravity, as well as other theory classes, it is possible to quotient out the reparametrizations that leave the physics invariant, leading to the so-called invariants of scalar-tensor gravity. In particular, the metric and Palatini formulations of scalar-tensor gravity and $F(R)$ gravity leave the corresponding actions form-invariant, without introducing any additional terms, unlike teleparallel formulations. The metric and Palatini formulation represents a discrete symmetry of the space of scalar-tensor gravity, as every model realized in metric has a corresponding, different model realized in Palatini that generates the same physics.

Motivated by this symmetry, we have introduced a way to interpolate between the metric and Palatini formulations with the aid of a single parameter that encodes the relative ``Palatininess'' of the model. This approach is distinct from hybrid metric-Palatini gravity, which interpolates between two distinct classes of (Brans--Dicke) actions. While the hybrid metric-Palatini actions are useful in motivating different models, they cannot be used to interpolate between two given formulations of a given, fixed model. There are two distinct ways to interpolate between different formulations: either via a linear combination of the gravitational sector scalars (the internal interpolation), or via a linear combinations of actions themselves (the external interpolation).

Given a model, the quasi-Palatini formulation leads to a continuous class of theories. We find that the internal and external interpolation prescriptions match for scalar-tensor models, but for $F(R)$ models, the external interpolation introduces an additional scalar degree of freedom.  We have studied the continuous class of distinct theories arising as a result of treating known models of inflation in the quasi-Palatini formulation, and we show that the resultant observables are enhanced with an additional parameter, increasing the robustness of these models.
    
Finally, we note that it is possible to interpolate between other formulations, giving rise to scalar-torsion and scalar-metricity theories as opposed to the simpler scalar-tensor theories tha arise as. By studying well-known models under this generalized formulation, the geometric trinity of gravity can be seen as the base of a prism: the base corresponds to GR, where the degrees of freedom are encoded in the curvature, torsion, or non-metricity, whereas the rest of the triangular slices of the prism correspond to different nonminimal models, in which the different formulations, a linear sum of scalar-curvature, scalar-torsion, and scalar-nonmetricity models, now correspond to different metric-affine theories. Studying well-established models in these quasi-formulations is expected to open up a fruitful avenue for model building, and to shed light on their astrophysical and cosmological implications. 

\subsection*{Acknowledgements} 
The author was supported by the Estonian Research Council Mobilitas 3.0 incoming postdoctoral grant MOB3JD1233 ``Inflationary Nonminimal Models: An Investigative Exploration''. The author would like to thank Laur J\"arv, Damianos Iosifidis, Margus Saal, María Jos\'e Guzm\'an Monsalve, and Benjamin Muntz for illuminating discussions and insightful comments.

\bibliographystyle{utphys} 
\bibliography{scalar_tensor_tidied}

%https://arxiv.org/pdf/1006.0454
%equivalence of metric/Palatini

%https://arxiv.org/pdf/0804.4440
%higher curvature gravity is equivalent metric/Palatini (for a class)

%https://arxiv.org/pdf/gr-qc/0403063
%all observable stuff is frame-invariant

\end{document}